\documentclass[10pt]{article}

\usepackage{amsmath}
\usepackage{xspace}
\input epsf.tex

\newcommand{\be}{\begin{equation}}
\newcommand{\bea}{\begin{align}}
\newcommand{\beano}{\begin{align*}}

\newcommand{\infinity}{{\infty}}
\newcommand{\angst}{\AA\xspace}
\newcommand{\bsigma}{\boldsymbol{\sigma}}
\newcommand{\grad}{\nabla}
\newcommand{\half}{ {\frac{1}{2}} }
\newcommand{\E}[1]{\!\times\!10^{#1}}

\newcommand{\vect}[1]{{\bf #1 }}
\newcommand{\subrm}[1]{\mbox{\scriptsize #1}}
\newcommand{\degree}{\ensuremath{^{\circ}}}
\newcommand{\rd}{{\,\mbox{d}}}
\newcommand{\bv}{\begin{verbatim}}
\newcommand{\refeq}[1]{equation (\ref{eq:#1})}
\newcommand{\reffig}[1]{Fig.\ \ref{fig:#1}}

\newcommand{\dea}{{ \Delta E^{a} }}
\newcommand{\deb}{{ \Delta E^{b} }}
\newcommand{\dec}{{ \Delta E^{c} }}
\newcommand{\ms}{{ \partial M_s }}
\newcommand{\msp}{{ \partial M_{s^{\prime}} }}
\newcommand{\me}{{ \partial M_e }}
\newcommand{\ims}{{ \int_{\partial M_s} }}
\newcommand{\imsp}{{ \int_{\partial M_{s^{\prime}}} }}
\newcommand{\ime}{{ \int_{\partial M_e} }}
\newcommand{\jup}{{ [[\vu_P]] }}
\newcommand{\juo}{{ [[\vu_O]] }}
\newcommand{\vu}{{ \vect{u} }}
\newcommand{\vt}{{ \vect{t} }}
\newcommand{\vb}{{ \vect{b} }}

\newcommand{\sumn}{{ \sum_{n=1}^{3} }}

\newcommand{\vuD}{{ \vu_{\Delta} }}
\newcommand{\vtD}{{ \vt_{\Delta} }}
\newcommand{\thp}{{ \theta^{\prime}}}

\newcommand{\PK}{Peach-Kohler\xspace}

\newcommand{\vs}{{ \boldsymbol{\sigma} }}
\newcommand{\ve}{{ \boldsymbol{\epsilon} }}

\renewcommand{\reffig}[1]{figure\ \ref{fig:#1}}

\begin{document}

\renewcommand{\thefootnote}{\fnsymbol{footnote}}

\title{Lattice Resistance and Peierls Stress in Finite-size
Atomistic Dislocation Simulations}
\author{David L. Olmsted \footnotemark[2]
        \and Kedar Y. Hardikar
        \and Rob Phillips \\
      Division of Engineering, Brown University
}
\maketitle
\footnotetext[2]{Corresponding author. E-mail address: olmsted@engin.brown.edu}
\renewcommand{\thefootnote}{\arabic{footnote}}

\begin{abstract}
{\normalsize
Atomistic computations of the 
Peierls stress in fcc metals are relatively scarce. 
By way of contrast, there are many more
atomistic computations for bcc metals, as well as mixed discrete-continuum
computations of the Peierls-Nabarro type for fcc metals. One of the
reasons for this is the low Peierls stresses in fcc metals. Because
atomistic computations of the Peierls stress take place in finite
simulation cells, image forces caused by boundaries must either
be relaxed or corrected for if system size independent
 results are to be obtained.
One of the approaches that has been developed for treating
such boundary forces is by computing them directly and subsequently
subtracting their effects,
as developed in Shenoy and Phillips \cite{shenoy97}. That work was
primarily analytic, and limited to screw dislocations and
special symmetric geometries. We extend that work to edge and mixed
dislocations, and to arbitrary two-dimensional geometries, 
through a numerical finite element computation. We also describe a 
method for estimating the boundary forces directly on the
basis of atomistic calculations.
We apply these methods to the numerical measurement of the Peierls stress
and lattice resistance curves for a model aluminum (fcc) system
using an embedded-atom potential \cite{ercolessi94}. 
} 
\end{abstract}

\subsection{Introduction}
Dislocation simulations at the atomistic level are affected by
boundary forces, except in very special cases. This is equally true
whether the simulations are based on semi-empirical potentials or
density functional theory. Two examples which serve to illustrate
the potentially disastrous influence of such boundary forces
are: (1) To estimate the Peierls stress for a given system, 
that is the applied stress needed to move a straight dislocation
in an otherwise perfect crystal. (2) To simulate the bow-out
of a pinned dislocation under an applied stress.

If one simulates a single dislocation in a finite cell,
then the boundary conditions imposed on that cell will
determine the nature of the resulting boundary forces. One 
common type of boundary
condition for a single dislocation is to fix all atoms whose distance
from the dislocation core exceeds some critical radius at their
positions as given by the linear (anisotropic) elastic
solution for the dislocation of interest. This assumes that the linear 
elastic solution is accurate
at long distances, which is a good assumption in many cases for a
straight undissociated dislocation at a fixed position.
To determine the Peierls stress the dislocation must be moved, however.
After the dislocation has moved, the locations of the atoms in the 
fixed region will not be consistent with the 
elastic field of the dislocation in its new position.

In such a simulation, the dislocation is moved in the cell
by applying a slowly increasing external stress. This external
applied shear stress is simulated by applying the
appropriate strain increment to all of the atoms, and, upon
relaxing the free region, the fixed positions of the atoms
in the exterior ring impose the desired stress on the free region.
The minimum applied stress necessary to jump the dislocation
out of its initial Peierls well, the apparent Peierls stress,
will be overestimated because not only must the applied stress overcome
the intrinsic lattice resistance, but the force due to the boundary
as well.

At least two methods have been used to obtain accurate estimates
of the Peierls stress of an atomistic model in cases where
the boundary forces are important. One method is to relax
the boundary forces through 
flexible boundary conditions \cite{rao98}.
The method we use here, developed by Shenoy and Phillips \cite{shenoy97},
[referred to hereafter as (I)],
is to determine the boundary force contribution and correct for it.
While this method is less elegant than the flexible boundary
condition approach, it does not require any changes to the
simulations themselves. We believe that this will make it
useful to researchers in many cases where either the 
importance of boundary forces have not yet been determined,
where implementing the flexible boundary condition is not
an efficient allocation of effort, or where a ``rough and ready''
approach is needed as a starting point.

The boundary force correction approach also has one 
valuable capability that flexible boundary conditions does not.
If the boundary forces are fully relaxed, the dislocation will
jump as soon as the Peierls stress is reached. This
creates a limitation on the part of the lattice resistance curve
that can be measured. Using fixed boundary conditions, although
the lattice resistance will begin to decrease after the Peierls stress
is reached, the boundary resistance is monotonically increasing.
The dislocation will not jump until the total resistance
begins to decrease. After subtracting the boundary force to
obtain the lattice resistance, some portion of the lattice
resistance curve beyond the area accessible to the flexible
boundary condition approach is maintained. In fact, if the
boundary forces are large enough, it is possible that the
entire lattice resistance curve will become accessible.
This is a useful, but imperfect benefit, because the small
cell required to view the entire lattice resistance curve is
likely to also be small enough to introduce distortions, even
after correction of the boundary forces. In the work reported 
here, only for the smallest cells is any data obtained for
the lattice resistance curve in the areas that would otherwise
be inaccessible.

A third possiblity which has been used for measuring the
static Peierls stress of fcc dislocations is fully periodic
boundary conditions \cite{bulatov99}. Since fully periodic boundary
conditions are inconsistent with the existence of a net
Burgers vector in the simulation cell, a dislocation dipole
or quadrupole must be used \cite{bulatov95}.
As discussed in ref. \cite{bulatov95} a disadvantage in this case
are the interactions between the dislocations and with their images.
In measuring the Peierls stress based on the motion of a dislocation in 
response to an applied stress, the dislocations of opposite Burgers 
vector making up the dipole or quadrupole will experience equal and 
opposite Peach-Kohler forces,
and their relative positions will change. While the long range
nature of the forces between the dipoles and images will complicate
any correction for the forces between the dislocations, these could
be computed in linear elasticity based on the measured locations of
the dislocations or partials. For a simulation cell of the same
size, the forces between the dislocations (including the periodic images)
will be stronger than the boundary forces in our configuration.
This increases the possibility of distortion of the dislocation core.
In the case of an fcc dislocation split into Shockley partials,
there will be a force compressing or expanding 
the partials which will be much stronger than
in the configuration used here. The extent to which the resulting
simulation size dependence of the partial separation will 
make the Peierls stress size dependent is unclear.  

A fourth possible approach to measuring the Peierls stress
would be to use periodic boundary conditions in the glide
direction as well as the line direction, but with walls of fixed 
(or partially fixed) atoms
on each side in the third direction.
We are aware of dynamic studies of dislocation motion
in such configurations \cite{daw86,daw93,rodney99a,rodney00}, 
but not of any measurements of the static
Peierls stress. Such a configuration can contain a single dislocation,
avoiding some of the hazards of the previous approach.
Compared to our approach, in such a configuration the forces generated
by the fixed boundaries will have no net component in the
glide direction. As in the case of fully periodic boundary conditions, 
for a simulation cell of similar size, the
forces between the periodic images of the dislocation will be
stronger than the boundary forces in our configuration, but
again the net force on the dislocation should be zero.
The possibilities of core distortion are again significant, however,
and similar to the case of full periodic boundary conditions.

In (I) the boundary forces are computed analytically for a screw
dislocation, and it is shown that this boundary force correction
allows for computation of a size-independent Peierls stress and
for the simulation of bow-out. The analytical solution presented is
limited to the screw dislocation, and is also limited to a circular geometry
for the case of isotropic elastic moduli. For anisotropic elastic
moduli it is limited to an elliptical geometry.
We have extended the computation of the boundary forces using the
approach developed in (I) in such a way that it can be used for both
dislocations of other characters and to other geometries. In fact,
we introduce two numerical schemes which allow for the determination of
the boundary force correction. One method is based upon
linear elasticity and is implemented numerically
using the finite-element method, and it provides a more
general numerical implementation of the ideas presented in (I).
The other method is a measurement of the energy associated with
the boundary-force directly on the basis of an atomistic description of
the total energy. 

We apply our methods to
measurements of the lattice resistance curves and Peierls stresses
for dislocations in aluminum, using the Ercolessi and Adams glue
potential \cite{ercolessi94}. The estimates of the lowest order term in
the boundary-force from the finite element method elasticity 
computations and from atomistics are in good agreement.
The computed boundary-force predictions allow estimates of the
Peierls stress that show much less size dependence than
computations done without such corrections. This makes it possible
for us to measure, with reasonable accuracy, the Peierls stress
for the easy-glide edge dislocation in this model system, which
is roughly $6 \E{-5} \mu$. In the absence of boundary
corrections, this value is overestimated by a factor varying
between four and two.
(The degree to which this stress is overestimated
 depends on the size of the simulation cell, 
and is off by a factor of four for a cell with a diameter of $100$ \angst,
and by a factor of two even for a cell with a diameter of $180$ \angst.)
The magnitude of the corrections vary with the dislocation
character. For the screw dislocation, where the Peierls stress
is roughly $5 \E{-4} \mu$, the overestimate caused by ignoring
boundary force corrections is significantly smaller than for the
edge dislocation.

\subsection{The boundary force on a dislocation.}

In order to correct for boundary force effects we must
understand them in quantitative terms and be able to compute them accurately.
Fortunately, for the case of an isotropic screw dislocation
in a cylindrical body, a closed form analytic result is available
from an image method [(I)]. Consider a screw dislocation
displaced a distance $d$ from the center of a cylindrical region
of radius $R$ as depicted in \reffig{cell_fig}.
We suppose that the boundary conditions imposed at the surface
are fixed displacements, corresponding to the isotropic linear
elastic solution for the dislocation located at the center.
We wish to obtain the excess energy of the system with the
dislocation located at $P$, where it is displaced by $d$, compared
to the system with the dislocation at the center, $O$.
To effect the calculation of this excess energy, we assume the
material can be characterized using isotropic linear elasticity.
\begin{figure}
\epsfxsize=\hsize \epsfbox{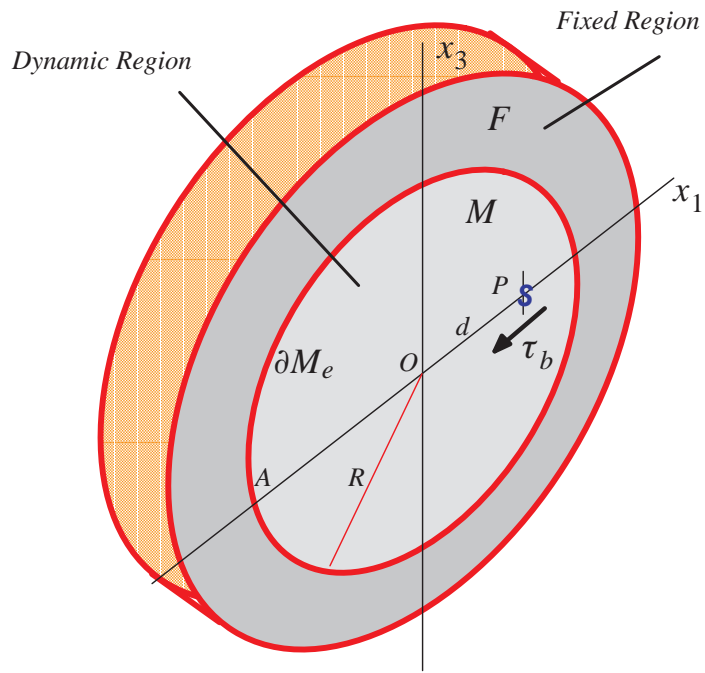}
\caption{\label{fig:cell_fig}
Cell geometry illustrating both the free and fixed regions (taken from
Ref. \protect \cite{shenoy97}).
}   
\end{figure}

In \reffig{image_fig} we illustrate the image solution.
By adding a 
screw dislocation at a point Q, with the same Burgers vector, it
is possible to create an infinite medium problem where the
displacements at the edge of the circle caused by the dislocations at P and
at Q satisfy the relevant boundary conditions which are the displacements for
a single dislocation at position $O$.
\begin{figure}
\epsfxsize=\hsize \epsfbox{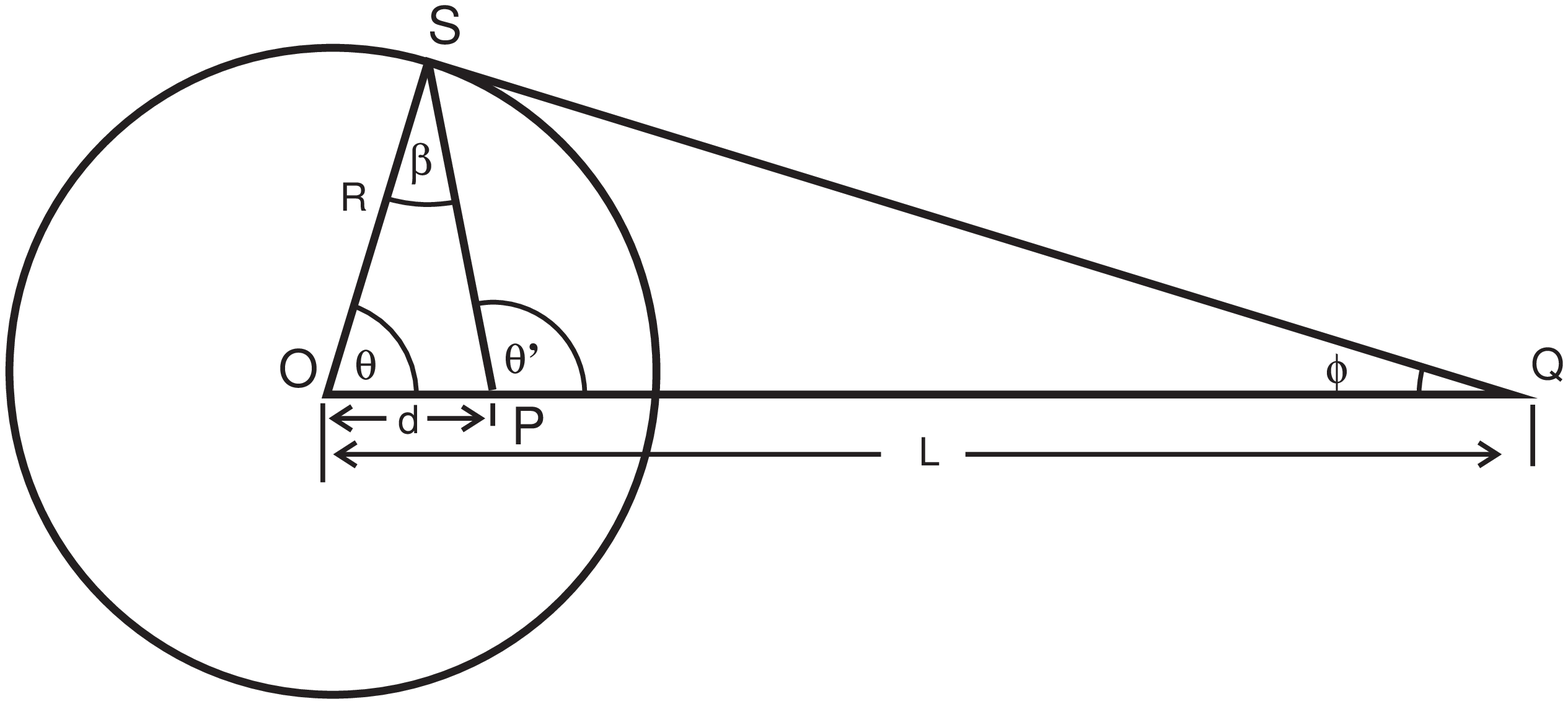}
\caption{\label{fig:image_fig}
Image solution
}   
\end{figure}

Consider a particular point on the circle, given by an angle $\theta$.
Let $\thp$ be the angle at $P$, and $\phi$ the angle at $Q$.
For the isotropic screw dislocation the displacements are entirely
in the $x_2$ direction, which is parallel to the dislocation
line itself, and perpendicular to the plane of the paper in 
\reffig{image_fig}. For the simple case of a screw dislocation,
the fixed displacement at point S due to a dislocation
at the origin is $-b \theta / 2 \pi$ \cite{hirth82}, and
we note that these are the displacements that serve as our boundary condition.
The displacement caused by a dislocation at $P$ is
$-b \thp / 2 \pi$. Taking the surface at which the displacement jump occurs 
for the dislocation at $Q$ to its right along the positive $x_1$ axis, 
the displacement it causes is $b \phi / 2 \pi$. 
From a physical perspective, we now require that the superposition 
of the fields due
to the dislocations at $P$ and $Q$ result in a displacement field 
that is entirely
equivalent to that due to the single dislocation at $O$. From a mathematical
perspective, this statement is 
\bea
  \thp - \phi &= \theta,      \\
  \phi &= \thp - \theta.
\end{align}
This determines the point $Q$ and the distance $L$, which at this 
stage depend on $\theta$. Notice that there are
further geometric constraints such as
\be
  \beta = \thp - \theta = \phi.
\end{equation}
The triangles $OQS$ and $OSP$ are therefore similar, and we have
\bea
  \frac{L}{R} &= \frac{R}{d}    \\
  L &= \frac{R^2}{d}.
\end{align}
Since $L$ is independent of $\theta$, the boundary condition
is satisfied at all  points on the circle. Because the dislocation
at $P$ feels no force from its own elastic strain field, the
total configurational force on it is the force caused by the dislocation
at $Q$, and is equal to \cite{hirth82}
\bea
  F_{\subrm{b}}/l &= - \frac{\mu b^2}{2 \pi} \frac{1}{L - d}  \\
      &= - \frac{\mu b^2}{2 \pi} \left( \frac{d}{R^2 - d^2} \right).
              \label{eq:image_solution}
\end{align}
We note that the image dislocation is at the same location, but
opposite in sign, to the image dislocation relevant to the same
cylindrical geometry, but with free-surface 
boundary conditions \cite{hirth82}.

This force profile can be used in turn to characterize the energetics
of the system as a result of the frozen boundaries. Indeed, 
integration over this force-displacement relation yields
 the extra elastic energy generated by displacing the
dislocation a distance $d$ from the center of the cylinder as
\be
  E/l = - \frac {\mu b^2}{4 \pi} \ln\left(1 - \frac{d^2}{R^2}\right).
\end{equation}
On the other hand, for dislocations other than the screw dislocation
considered above or for anisotropic elasticity,
we do not expect such a simple, closed-form solution. Noting, however,
that for this particular solution the energy depends only on $d/R$, not on 
$d$ and $R$ separately, and has a convergent power series 
expansion in the relevant range, $-1 < d/R < 1$, in our
subsequent developments we have been emboldened to
fit our numerical results to a power series in $d/R$. 

For our model aluminum system, the screw dislocation splits 
(approximately) into Shockley partials. While we cannot use the image
solution to handle the edge portions of the partials, it can be
used to compute the boundary force for the non-physical case of 
pure screw partials, each with Burgers vector equal to
$b/2$. This is discussed in Appendix A. (In general
all of the treatment of the affects of partial splitting have
been placed in Appendix A, in hopes of making the story in the main
text more focused.)

The use of \refeq{image_solution} to determining the lattice
resistance curve was developed in (I). There it was assumed that
total configurational force on the displaced dislocation has three
contributions, the \PK force corresponding to the applied
shear stress, which is independent of $d$; the boundary force,
which in linear elasticity (for a perfect dislocation) 
depends only on $d/R$; and
the lattice resistance, which is assumed to depend 
periodically on $d$, and to be independent of $R$.
As discussed in Appendix A, the splitting of the dislocations
into Shockley partials allows the boundary force to depend on
both $d/R$, and $s/R$, where $s$ is half the distance between
the partials. As in (I) we assume that the partial separation
is fixed during the course of the dislocation's journey across the
Peierls energy landscape. 

For a configuration of the system where the dislocation is at
static equilibrium, we then have 
\be
    F_{\subrm{app}} + F_{\subrm{b}}(d) + F_{\subrm{L}}(d) = 0, 
      \hspace{3em}
    \label{eq:force_balance}
\end{equation}
The applied force is given by the \PK formula. If, for a given
applied force, we can measure the displacement $d$ of the
dislocation from the center of the cell, and if we can
compute the boundary force, then subtracting the boundary
force we obtain the value of the lattice resistance,
$F_{\subrm{L}}(d)$. (Our measurement of the position of the
dislocation is discussed below.)
By varying the applied stress we obtain the curve $F_{\subrm{L}}(d)$,
which allows us to determine not only the Peierls stress, but to map
out the Peierls force landscape as well.

An alternative way to view $F_{\subrm{b}}(d)$ is to say that 
the elastic stress to be used in the \PK formula is the applied
stress less the amount of stress relieved through 
the movement of the dislocation \cite{simmons97,rao98}.
Using this approach Simmons {\it et al.} \cite{simmons97}
 were able
to obtain the correct scaling for the leading term in $F_{\subrm{b}}(d)$,
that is $\mu b d/R^2$, and a reasonable estimate of its numeric
coefficient (which did not vary with dislocation character).

Before embarking on a precise numerical assessment of our results,
we first need to consider the validity of the assumptions
leading to \refeq{force_balance}.
As the dislocation is displaced with respect to the
periodic lattice, the core configuration will change. Because
the configurational forces generated by the applied stress and
the boundary conditions are both long-range elastic effects, we
expect the assumption that they are unaffected by this periodic
change in the core structure to be negligible for
$d \ll R$. It is possible that the core configuration will
be affected by the boundary forces, however, and that this
will cause some distortion of the lattice resistance force
as the dislocation moves. As discussed in (I) this is certainly
possible for the partial separation distance, since the
boundary forces on the two partials will tend to have a net
constrictive element. Clearly \refeq{force_balance} is an
assumption we need to validate for our system.

As described above, our results will provide the curve
$F_{\subrm{L}}(d)$.
By assumption this will depend periodically on $d$ and be
independent of $R$. By testing these predictions
in our results, we will find that the underlying assumptions
are appropriate for all of our dislocations except the edge,
and are reasonable for the edge dislocation, 
except for our smallest cell.

\subsection{Computation of Boundary Force Correction Coefficients}

\subsubsection{\em Setup}

As in (I) we consider the problem of modeling a straight
dislocation in a finite cylindrical cell. The geometry, shown in
\reffig{cell_fig}, has $x_2$ as the line direction of
the dislocation, and the $x_1$--$x_2$ plane as the slip plane.
In the simulation there is a dynamic region of radius R, and
a fixed region outside R, with periodic boundary conditions 
in the $x_2$ direction. A dislocation is introduced
at $O$, the center of the cell, using the continuum anisotropic 
linear elastic solution for either a single dislocation or for
two Shockley partials. This initial configuration is relaxed
using energy minimization, by allowing the atoms in the free region
to move. The atoms in the fixed region remain at the locations
corresponding to the elasticity solution.
This relaxed dislocation at the
center of the cell serves as a reference configuration.

The objective of our analysis is to examine the response of a
dislocation to an applied stress, while correcting for the 
contaminating influence of boundaries.
We model the application of a homogeneous shear stress by
moving the atoms according to the homogeneous strain
which produces the desired stress under linear elasticity.
Keeping the atoms in the fixed region at these strained positions,
the atoms in the dynamic region are relaxed. The dislocation now
moves to a new position $P$ which (locally) minimizes its energy. 
We assume that the structure of the dislocation core does not change, 
and we therefore treat the distance $d = \overline{OP}$ as 
the sole configurational parameter. 

Consider, then, the energy per unit length, $E(d)$, of the
dislocation as a function of its
distance $d$ from the center of the cylindrical cell. 
As in (I) we assume that this configurational energy consists of
four parts
\be
     E(d) = E_{\subrm{ref}}
            - d b \tau_{\subrm{app}}
            + E_{\subrm{L}}(d)
            + E_{\subrm{b}}(d),
                            \label{eq:totalEnergy}
\end{equation}
where $E_{\subrm{ref}}$ is a constant representing the 
energy of the reference configuration when the dislocation is at
the origin;
$\tau_{\subrm{app}}$ is the resolved shear stress;
$E_{\subrm{L}}$ is the Peierls energy, assumed to be periodic
in $d$; and $E_{\subrm{b}}$ is the energy associated with the
boundary force. The first three terms appear in the energy
of a dislocation in an infinite medium, while $E_{\subrm{b}}$
is the result of the mismatch between the imposed displacement boundary
conditions and the strain field the dislocation would generate at
its position $d$ in an infinite medium. 
And, as in (I), we define 
\be
    \Delta E = E_{\subrm{b}}(d) - E_{\subrm{b}}(0).
\end{equation}

Motivated by the analytic structure of the image solution discussed
above,
we will couch our estimates for the energy 
caused by the finite boundaries (for small $d/R$) in the
form
\be 
    \Delta E = \frac{\mu b^2}{8 \pi^2} \left[
       A \left( \frac{d}{R} \right)^2
      + C \left( \frac{d}{R} \right)^3
      + B \left( \frac{d}{R} \right)^4
      + O((d/R)^5) \right].
	\label{eq:PowersEquation}
\end{equation}
Here $\mu$ is taken as $\{C_{44}^2 + [(C_{11}-C_{12})/2]^2\}^{1/2}$,
the value of $\mu$ which gives the correct logarithmic prefactor for
the line energy of the (anisotropic) screw dislocation.
By symmetry, odd powers can only occur for the case in which
a mixed-character dislocation is treated as split into
Shockley partials (as the partials then have different characters).
For the isotropic screw dislocation, the image solution then gives
$A = 2 \pi; B = \pi$. 

In the application of this expansion in powers of $d/R$ to the
simulations there is some ambiguity in the measurement of both
$d$ and $R$. The measurement of $d$ is discussed below. In the
analytic elasticity solution using the image solution, the fixed displacement
boundary conditions are specified at all points on the circle of radius $R$.
In the simulations, the fixed atoms are necessarily discrete. 
The specification of fixed displacement boundary conditions at all the
atomic positions which lie outside a circle of radius $R$ provides a `softer'
boundary condition than if all points on the circle of radius $R$ had
specified displacements, making the effective $R$ that best fits the expansion
slightly greater than the $R$ used to specify the fixed region. 
Except as otherwise
stated, in this work we have used the nominal $R$ for the simulations.
(It is also the case that the nominal $R$ is 
ambiguous in that it could equally well
be any value between the value of $r$ for the outermost free atom
 and the value of $r$ for the innermost fixed atom. 
This difference is very small, and is small
compared to the difference between the `nominal' and the `effective' $R$.

\subsubsection{\em Computing Coefficients in Linear Elasticity} 

In order to correct for the finite system size by using
the force balance equation, \refeq{force_balance}, 
we need to evaluate the boundary force,
which we derive from the configurational energy $\Delta E$.
We estimate $\Delta E$ in this section in linear elasticity,
using a finite element method numerical solution. In the following
section we discuss a second approximate scheme for the
determination of $\Delta E$ using atomistic simulations.

The evaluation of $\Delta E$ in linear elasticity follows (I) 
exactly, except for the method of computation. The energy
difference $\Delta E$ is the excess of the elastic 
energy $E_2$ of the system with the dislocation
at $P$, over its energy $E_1$ with the dislocation at $O$.
The boundary conditions at the edge of the finite cylinder are
fixed displacements, given by the linear elastic (Volterra) solution
for the dislocation at $O$. This element of the boundary conditions
is the same whether the dislocation is at $O$ or at $P$. The
boundary conditions on the slip surface are given by 
$[[\vu]] = \vb$, but the slip surface is extended from
$O$ to $P$ when the dislocation moves to $P$. (Or retracted from
$O$ to $P$ if $d<0$.) This element of the boundary conditions, therefore, 
differs by a displacement jump between the two configurations.
We know the stress, strain and
displacement fields when the dislocation is at $O$ analytically, from
the sextic formulation of Eshelby et al. 
\cite{eshelby53,foreman55,teutonico63,sextic}.
The energy for the system at $O$ is then simply
\be
E_1 = \half \int_{M} \vs_O \cdot \ve_O \rd V,
\label{eq:vol_integral}    
\end{equation}
where elastic variables with a subscript $O$ refer to the
elastic fields of a dislocation at $O$ in an infinite medium.
As written, $E_1$ has the normal logarithmic divergence at $O$.
We handle this in the customary way, excluding a cylinder of radius
$r_c$ at $O$, and introducing a dependence of $E_1$ on $r_c$ \cite{hirth82}.
As in (I), we assume the same value of $r_c$ in evaluating $E_1$ and
$E_2$, in which case the dependence on $r_c$ cancels in taking
the difference $E_2 - E_1$, which is independent of $r_0$, as
will be seen below.
 
When the dislocation is at $P$ we do not have analytic solutions for
the displacement, stress and strain fields available. 
We wish to apply the finite element method to numerically solve
the elasticity problem posed by the difference between the
two configurations with the dislocation at $O$ and $P$. However,
rather than attempt to handle the displacement jump boundary
condition in the finite element method, it is convenient to 
separate the problem into two problems, one of which has 
only (continuous) fixed displacement boundary conditions, and
one of which has analytic solutions for the fields available.
We therefore also consider a third set of elastic fields, those
for a dislocation at $P$ in an infinite medium. (They will
be denoted with a subscript $P$, while the equilibrium fields for the
configuration with the dislocation at $P$ subject to our
actual boundary conditions will have no subscript.)
We therefore define 
\be
 \vu_{\Delta} = \vu - \vu_P.
\end{equation}
The interpretation of this is 
that the displacement field due
to the dislocation when it is at $P$ is that of a dislocation
situated at $P$ in an infinite body, 
plus a correction field $\vuD$ which
adjusts $\vu_P$ so as to be consonant with the boundary conditions.
The boundary conditions on $\vu$ are
\bea
 \vu =& \vu_O    \hspace{3em} \mbox{ on } \me \notag \\
 [[\vu]] =& \vb  \hspace{3em} \mbox{ on } \msp.
\end{align}
Since $\vu_P$ also has $[[\vu_P]] = \vb$ on $\msp$, the boundary
conditions for $\vuD$ are
\bea
 \vuD =& \vu_O - \vu_P    \hspace{3em} \mbox{ on } \me \notag \\
 [[ \vuD ]] =& \vect{0}  \hspace{3em} \mbox{ on } \msp.
\end{align}
Thus $\vuD$ is the solution to an elasticity boundary value
problem without singularities, and can be solved by the
finite element method.
The fields $\vu_P$, and the related stress and strain fields
are essentially the same as the fields $\vu_0$ save that they have been translated.

We now have
\be
  E_2 = \half \int_{M} (\vs_P + \vs_{\Delta}) 
                 \cdot (\ve_P + \ve_{\Delta}) \rd V,
\end{equation}
where
\bea
   \ve_{\Delta} =& \grad \vu_{\Delta} \notag \\
   \vs_{\Delta} =& \vect{C} : \ve_{\Delta},
\end{align}
where \vect{C} is the elastic stiffness  tensor.

In principle, these are satisfactory forms in which to
compute the energy difference. However, for purposes of numerical
computation it is preferable to manipulate $E_1 - E_2$ into
a form where the portions not involving $\vuD$ are reduced from
volume (effectively surface) integrals to an analytic piece plus
a line integral. Among other advantages, this eliminates
the logarithmic divergences at $O$ and $P$, as the cancellation
is handled in the analytic piece.
The rest of this section, which follows (I),
develops this form of $E_1 - E_2$.
 
The energy of the dislocation in the finite cylinder can be
computed as the work done by applying tractions on all relevant
surfaces to obtain the final displacements, 
where these surfaces must include
not only the exterior surface of the cylinder, but also a slip surface
where the displacement discontinuity of the dislocation is introduced.
We choose the plane $\overline{AP}$, denoted $\msp$ 
($\overline{AO}$, denoted $\ms$) for the slip surface when the
dislocation is at $P$ ($O$, respectively).
We denote the exterior of the cylinder as $\me$. Notice that, except
for the case of a screw dislocation, the surface of the cylinder will have
a step at $A$ either before or after the dislocation is inserted. 
This step is
the same in both configurations, however, and it will not contribute
to $\Delta E$. For convenience in numerical computation, we assume
that the cylinder started with a step at $A$, which is eliminated
by the insertion of the dislocation.
The elastic energy of the system with the dislocation
at $O$ is given by [(I)],
\be
  E_1 = \half \ims \vt_O \cdot \juo \rd S
        + \half \ime \vt_O \cdot \vu_O \rd S,
\end{equation}
where $\vt_O$ is the traction at the surface caused by the
presence of the dislocation at $O$ {\em in an infinite medium}, 
$\vu_O$ is the displacement at the exterior surface, and
$\juo$ is the displacement jump at the slip plane.
(We have $\vt_O = \bsigma_O \cdot \vect{n}$, where
$\bsigma_O$ is the stress tensor, and $\vect{n}$ the outward
normal for the exterior surface and is in the positive
$x_3$ direction at the slip plane.) $\juo$ is the jump in
displacement experienced on crossing the slip plane in the opposite
direction to $\vect{n}$, and is equal to $\vb$, the Burgers vector.
The line direction here is in the positive $x_2$ direction, and
our choice of directions is that of (I).

Letting $\vt$ be the traction associated with $\vu$ and
$\vtD$ be the traction associated with $\vuD$, we
have by linearity that
\be
 \vtD = \vt - \vt_P.
\end{equation}
The energy of the configuration with the dislocation at $P$ is now
\bea
  E_2 =& \half \imsp \vt \cdot \jup \rd S
        + \half \ime \vt \cdot \vu_O \rd S        \\
      =& \half \imsp (\vt_P + \vtD) \cdot \jup \rd S
        + \half \ime (\vt_P + \vtD) \cdot (\vu_P + \vuD) \rd S.
\end{align}
By the reciprocal theorem
\begin{eqnarray}
\lefteqn{  \half \imsp \vtD \cdot \jup \rd S
   + \half \ime \vtD \cdot \vu_P \rd S }  \hspace*{5em}    \notag \\ 
  &=& \half \imsp \vt_P \cdot [[\vuD]] \rd S
   + \half \ime \vt_P \cdot \vuD \rd S        \notag \\
  &=& \half \ime \vt_P \cdot \vuD \rd S.
\end{eqnarray}
Hence, following (I) we have,
\bea
  E_2  =&  \half \imsp \vt_P \cdot  \jup  \rd S
        + \half \ime  \vt_P \cdot  \vu_P \rd S     \notag \\  
       & \mbox{} + \ime  \vt_P \cdot  \vuD   \rd S
        + \half \ime  \vtD   \cdot  \vuD   \rd S.
\end{align}

For computational purposes, we split
$\Delta E = E_2 - E_1$ into three parts.
\be
    \Delta E = \dea + \deb + \dec,
\end{equation}
 where
\bea
    \dea =& \half \imsp \vt_P \cdot \jup \rd S
           - \half \ims \vt_O \cdot \juo \rd S,      \\
    \deb =& \ime \vt_P \cdot ( \vu_O - \half \vu_P) \rd S
            - \half \ime \vt_O \cdot \vu_O \rd S,     \\
    \dec =& \half \ime \vtD \cdot \vuD \rd S.
\end{align}

Of the three parts of this energy change, $\dea$ involves only the same terms
used to compute the energy of a dislocation in an infinite medium.
We take $r_c$ as a core cutoff radius, and K as the factor
defined by Hirth \& Lothe \cite{hirth82} eq. (13-83) 
for anisotropic elasticity. In isotropic elasticity
\be
   K = \mu \cos^2(\theta) + \frac{\mu}{1-\nu} \sin^2(\theta),
\end{equation}
where
$\theta$ is the angle the Burgers vector makes with the line
direction.
We then have, again following (I),
\bea 
  \frac{1}{l} \half \ims \vt_O \cdot \juo \rd S
    =& \frac{K b^2}{4 \pi} \int_{-R}^{-r_c} \frac{1}{-x_1} dx_1 \notag \\ 
    =& \frac{K b^2}{4 \pi} \ln\left(\frac{R}{r_c}\right),      \\
  \frac{1}{l} \half \imsp \vt_P \cdot \jup \rd S
    =& \frac{K b^2}{4 \pi} \int_{-R}^{d-r_c} \frac{1}{d-x_1} dx_1 \notag \\ 
    =& \frac{K b^2}{4 \pi} \ln\left(\frac{R+d}{r_c}\right),      \\
  \dea/l =& \frac{K b^2}{4 \pi} \ln\left(\frac{R+d}{R}\right).
        \label{eq:Ea}
\end{align}
The reader is encouraged to examine appendix A for the more complex 
case 
of a dislocation split into partials.

The second piece of $\Delta E$,
\be
    \deb = \ime \vt_P \cdot ( \vu_O - \half \vu_P) \rd S
            - \half \ime \vt_O \cdot \vu_O \rd S,
                 \label{eq:Eb}
\end{equation}
involves only the known elastic fields for anisotropic
dislocations in infinite media, evaluated on the boundary
surface. We therefore evaluate $\deb/l$ directly as
a numeric line integral around the circle.

The third piece of $\Delta E$ can be converted to
a (effectively two-dimensional) volume integral
\bea
    \dec =& \half \ime \vtD \cdot \vuD \rd S  \\
          =& \half \int_{M} \bsigma_{\Delta} : \grad \vuD   dV  \\
        =& \half \int_{M} \grad \vuD : \vect{C} : \grad \vuD dV,
             \label{eq:Ec}
\end{align}
where \vect{C} is the elastic stiffness  tensor.	  
We compute $\vuD$ as the solution to the relevant boundary value
problem described above, using the finite element method, and
so obtain $\dec/l$. 

This treatment extends the implementation of the approach
developed in (I) for the computation of the boundary force to
edge and mixed dislocations, and will handle arbitrary two-dimensional
geometries. The applications reported here are all for circular
geometries, though we have also applied the method to rectangular
geometries.

\subsubsection{\em Computing the Quadratic Coefficient from Atomistics}
The discussion given thus far has emphasized the use of the linear theory
of elasticity to evaluate the force on a dislocation as a result of 
the boundary conditions to which the system has been subjected.
However, as noted in the opening discussion of this paper, it is also 
possible to evaluate these boundary condition induced forces directly on
the basis of atomistic calculations, and that is the charter of the present
discussion.
As in the elasticity computation, an energy approach to computing
the configurational force is adopted. A dislocation is introduced
into the simulation cell using the (anisotropic) linear elastic
solution. The atoms in the exterior region are fixed at this solution,
and the dislocation is relaxed. The potential energy function 
is then interrogated
as to the total energy of the system, including that of the
fixed atoms. This will be referred to as the energy of the reference 
configuration. This energy includes a large surface energy contribution,
but it is a fixed value and will drop out of the differences in
energy computed as the dislocation is moved.

The dislocation is then moved by a lattice vector, so that, if the
lattice were infinite, the new configuration would have identical energy.
To measure the boundary-condition contribution to the energy, we
need to relax the free region of the simulation cell, subject to the
condition that the dislocation stay at its new location. For mobile
dislocations, if we relax all of the free atoms, the dislocation will
move back to the center of the cell. We therefore freeze a small
cylindrical region about the new location of the dislocation.
Because the relaxed dislocation cores are extended in the slip plane, this
frozen-core is chosen large enough to include the apparent
centers of both partials. For all the simulations used to compute
the boundary-force coefficient $A$ that was introduced in
\refeq{PowersEquation}, 
the free region had a radius of 127 \angst, and the fixed
solution for the exterior region was the (anisotropic) linear
elastic solution for a perfect dislocation. 
For a single value of the frozen-core radius the dislocation was
moved in the slip direction by at least five different distances
$d$ (in each direction) and the remaining free region relaxed.
The excess of the energy over the reference configuration was then
fit as a polynomial in $d/R$ giving an estimate of $A$. This
estimate now depends on the frozen-core radius, $r_{\subrm{fc}}$. We 
therefore repeated this process for at least three values 
of $r_{\subrm{fc}}$ for each dislocation character, and extrapolated 
to $r_{\subrm{fc}}=0$ as an estimate of $A$ for the dislocation
without a frozen-core.
In performing this extrapolation we first considered fitting 
to $r_{\subrm{fc}}^2$ since the volume of the frozen-core
scales with this parameter. 
The validity of this extrapolation scheme can be tested in
turn by appealing to the case of the Lomer dislocation which is
sessile and thus does not require the use of such a frozen-core
region. 
(The Lomer dislocation is a dislocation with a Burgers vector of
$a_0/2$ [1 -1 0] and a line direction of [1 1 0].)
Indeed, the Lomer case supports the hypothesized $r_{\subrm{fc}}^2$ 
scaling and hence
that has been used in the case of the glissile dislocations. 
\reffig{frozen_core} shows the quadratic fits to $r_{\subrm{fc}}^2$ used to
develop the atomistic estimates of A. A similar attempt to measure
B by extrapolation from frozen-core simulations did not produce
useful estimates of B.

\begin{figure}
\epsfxsize=\hsize \epsfbox{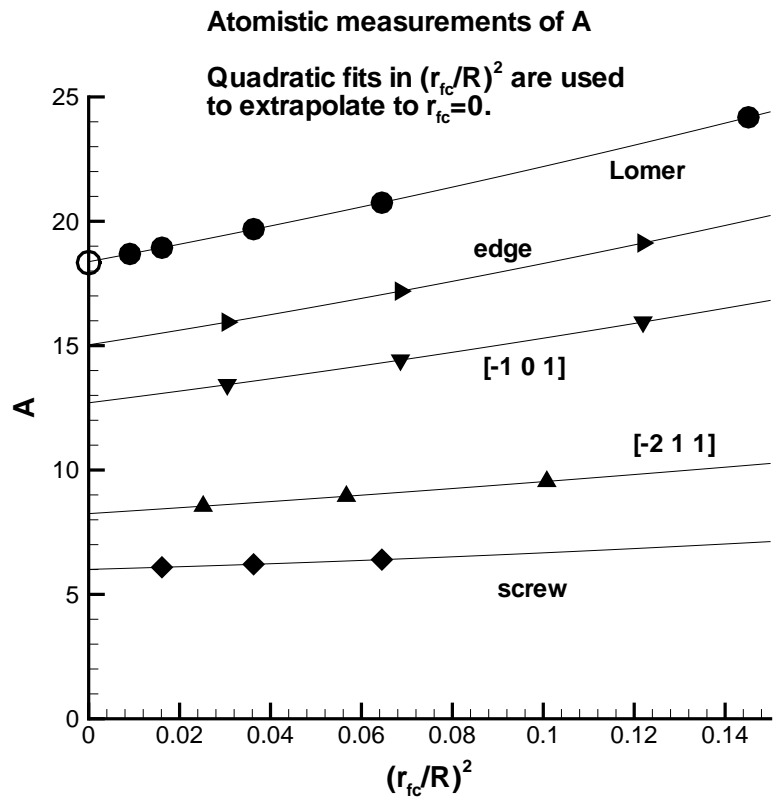}
\caption{\label{fig:frozen_core}
Extrapolation of the quadratic energy coefficient A as measured in
frozen-core simulations to $r_{\subrm{fc}}=0$. The extrapolations are
quadratic in $r_{\subrm{fc}}/R$. For the Lomer dislocation the open
circle is from the simulations without any frozen-core.
}   
\end{figure}

\reffig{comp_graph} shows the values of the coefficient $A$ of
the quadratic term in the boundary-force energy
for various dislocations as obtained using finite element computations, 
showing the difference between isotropic and anisotropic elasticity, 
and from the atomistic measurement. There is a significant difference 
in the boundary force coefficient
between the (glissile) edge dislocation and the Lomer dislocation, with
A differing by 20\% when anisotropic linear elasticity is used.
Given the close approximation of aluminum to isotropic elastic
constants this is rather large.
By comparison, the long-range portion of the dislocation
line energy differs by only 2.5 \% between the Lomer dislocation and the 
glissile edge dislocation. 
We also note that the atomistic values for `A' are slightly less 
than the values
from anisotropic elasticity. One cause for this discrepancy is
the `softer' boundary condition of the atomistic simulations.
On the other hand, given the severe differences in the calculational 
philosophy behind our two schemes [i.e. i) elasticity using finite 
elements and ii) direct atomistic evaluation of the boundary force] 
the level of agreement between the two schemes is remarkable.

\begin{figure}
\epsfxsize=\hsize \epsfbox{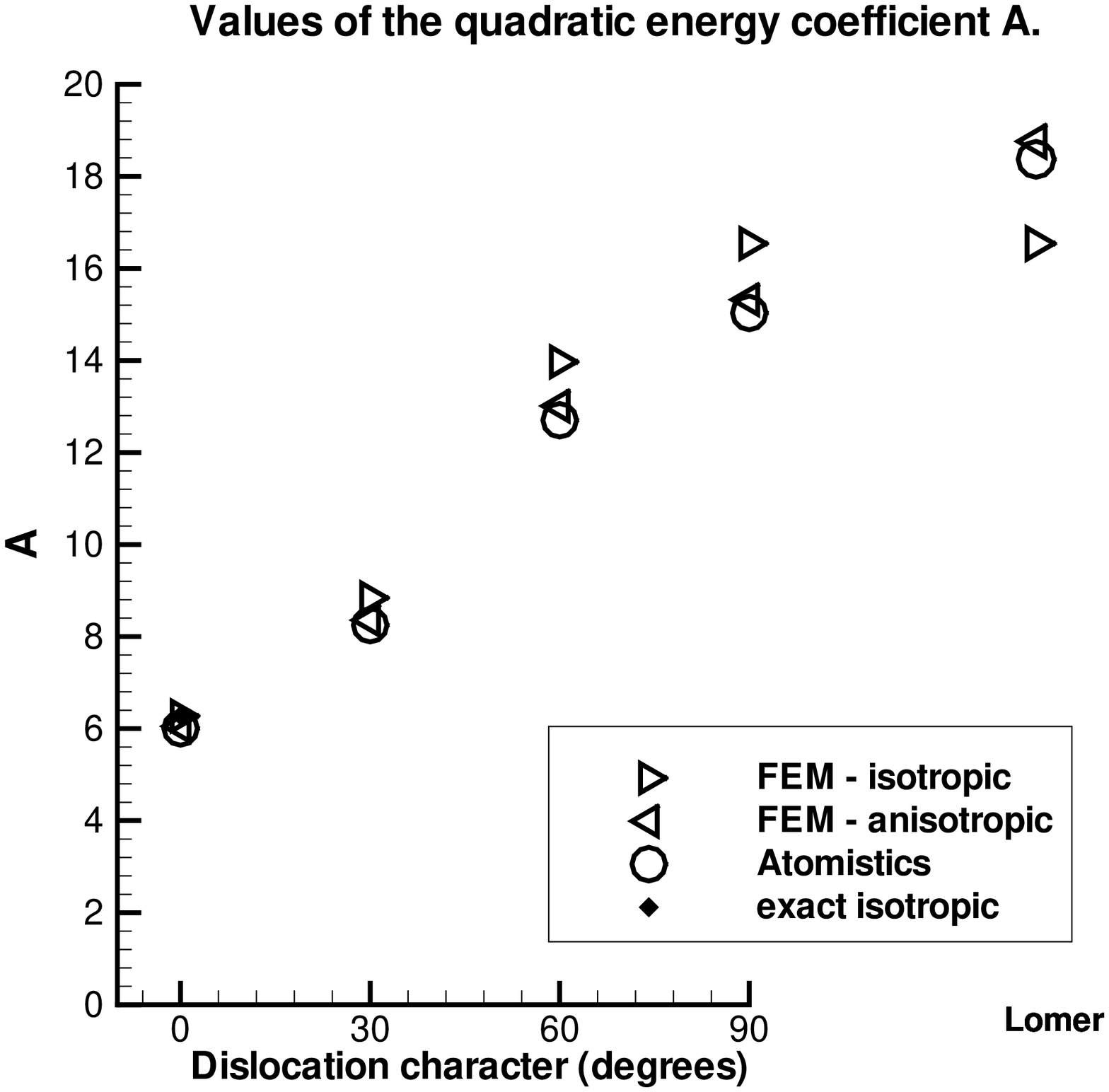}
\caption{\label{fig:comp_graph}
Quadratic coefficient in the boundary-force energy. This figure
shows the different approximations to the quadratic boundary force 
coefficient including isotropic elasticity, anisotropic elasticity, 
and direct atomistic calculation. The results are shown for five 
$a_0/2 [1 \bar{1} 0]$ dislocations; four are in the $(1 1 1)$ slip plane, 
labeled by the angle between the line direction
and the Burgers vector, with the screw dislocation labeled 0 and the
edge dislocation labeled 90. A sessile edge dislocation, the Lomer
dislocation, is also shown.
}   
\end{figure}

\subsection{Application of Boundary Force Correction}
The usefulness of the boundary force corrections proposed in
(I) depends on an accurate computation of the boundary force
 acting on a dislocation.
To compute the Peierls stress and the shape of the lattice 
resistance curve, a knowledge of the boundary force is
required only for small excursions of the dislocation from
its host well. On the other hand, to compute
the relevant forces in cases such as bow-out, it is
necessary that the boundary force be known
for large excursions of the dislocation from its
host well. As shown
in (I), if the computation is accurate for a straight dislocation,
good results can be achieved even in the case of bow-out by using a local
approximation based on the straight-dislocation results. As a result,
we emphasize the analysis of boundary forces on straight dislocations.

In an infinite medium, as the dislocation moves through the crystal
its energy (as a function of its displacement in the slip
plane perpendicular to its line direction) is a periodic
function of displacement \cite{hirth82}. We will refer to this as the Peierls
energy landscape. 
For the finite cell, with
fixed displacement boundary conditions, the dislocation is repelled
by the boundaries. The position that the dislocation will
move to under a given applied stress now depends on the
boundary force, the lattice resistance, and its starting position.
However, from \refeq{totalEnergy} we have that
at equilibrium [(I)]
\begin{gather}
    F_{\subrm{app}} + F_{\subrm{b}}(d) + F_{\subrm{L}}(d) = 0, 
      \hspace{3em}
    \label{eq:force_balance_2}             \\
\bea
    F_{\subrm{app}} &=  b \tau_{\subrm{app}},        \\
    F_{\subrm{b}} &= -\frac{\partial E_{\subrm{b}}}{\partial d},  \\
    F_{\subrm{L}} &= -\frac{\partial E_{\subrm{L}}}{\partial d}.
\end{align}
\end{gather}
It is possible to test the assumptions leading to
\refeq{force_balance} and our estimate of the
boundary force by computing the lattice resistance force as a function
of the dislocation's displacement from the relevant well.
If the boundary force terms have been handled correctly, 
and those assumptions hold,
the resulting
lattice resistance curves should be periodic in $d$ and should exhibit
an indifference to the size of the computational cell as characterized
by the parameter $R$.

\subsubsection{\em Simulations}
Like the calculations described above, 
the simulations used to test the boundary force correction
considered aluminum using the Ercolessi and Adams glue potential.
To test the dependence on the radius of the free portion of the  
simulation cell, $R$, we performed simulations for
$R=50$, 70 and 90 \angst. In addition, for the screw dislocation,
$R=30$ \angst was added, as discussed below. 
The dislocations considered had line
directions of [-1 1 0] (screw), [-2 1 1] (30$\degree$),
[-1 0 1] ($60\degree$), and [-1 -1 2] (edge); with a Burgers
vector of $a_0/2$ [1 -1 0].

We move the dislocation by creating a shear stress within the
simulation cell. We generate the shear stress by first applying
a uniform strain, corresponding to the shear stress to be applied,
to all of the atomic positions. The atoms in the free region
are then allowed to relax their positions, but with the atoms in the
fixed ring held at their strained positions.
The choice of applied stress is complicated by the splitting of the
perfect dislocation into Shockley partials with unequal
Burgers vectors. (And the conversion of the chosen applied stress
to an applied strain is complicated by anisotropy; see, for example,
M.S. Duesbury \cite{duesbury89}.) 
It is clearly desirable to choose the applied stress so that
the glide forces on the two partials are equal. Otherwise
the applied stress will tend to modify the dislocation core by
changing the partial separation. 
We chose the applied stress to provide the same force on
each nominal Shockley partial of the dislocation, including both
the glide and non-glide components. For the
edge and screw dislocations it was possible to do this while
keeping the non-glide component of the force zero. 
For the mixed dislocations there
was an applied climb force, which is not expected to
significantly affect the results.

The linear elastic solution used was that for the
two nominal Shockley partial dislocations corresponding to the
perfect dislocation of interest. These were 
assumed to be equidistant from the center
of the cell, with a separation based on the approximate
separation measured for a relaxed dislocation in earlier simulations.
The long-range elastic forces generated by the boundaries encourage
the dislocation to sit at the center of the simulation cell, 
in the absence of
any applied stress. A measurement of the Peierls stress that ignores
boundary forces in this geometry will depend on how far the dislocation
is from the center of the disk when it jumps from one well to the next, and
therefore will depend on how far the center of the Peierls well is from
the center of the disk. This will also have a smaller affect on
our corrected estimate of the Peierls stress. To give a `fair' representation
of the uncorrected estimate, we have aligned the simulation cell so that
the dislocation is near the center of a Peierls well when it is at
the center of the disk.

\subsubsection{\em Measurement of the position of the dislocation}
Note that in order to apply the boundary force correction, it is
necessary to know how large an excursion $d$ the dislocation
has taken from the origin. This, in turn, demands that we have
a scheme for identifying the position of the dislocation.
To effect this estimate we exploit the fact that a dislocation
is characterized by a jump discontinuity in the displacement fields
across the slip plane.
In particular, we define
$[[\vu(x)]]=\vu_{+}(x)-\vu_{-}(x)$, where $\vu_{+}(x)=\vu(x,0^+,z)$ 
is the displacement on one side of the slip plane, 
and $\vu_{-}(x)=\vu(x,0^-,z)$ is the displacement on the other side. 
(Here z is arbitrary, because of translation invariance in the 
line direction.) 
A difficulty that arises in making the transcription between continuum
notions such as that of the displacement field and a set of atomic 
positions is that quantities such as
$\vu(x,y,z)$ are actually defined only at the atomic sites.
We estimate $[[\vu(x)]]$ following a procedure of 
R. Miller and R. Phillips \cite{miller96}. 
$\vu_+(x)$ is estimated, for each x corresponding to an atomic
position in the planes of atoms nearest the slip plane, by looking at the
atoms in the first two planes on the `+' side of the slip plane. Given the
fcc crystallography, the atoms in the second plane do not lie above the atoms
in the first plane. However the projection of the atom in the first plane
onto the second plane lies at the center of an equilateral triangle of atoms
(in the perfect crystal). The average displacement of these three atoms
 in the second plane allows us to
define the displacement field in this plane as well.
Once the displacements in these two planes are in hand,
they are used to linearly extrapolate to $y=0$ in order
to obtain an 
 estimate of $\vu_+(x)$. Similarly $\vu_-(x)$ is estimated based on the first
two planes of atoms on that side of the slip plane.

The value of the displacement jump, $[[\vu(x)]]$, for large $x$, and the value 
for small $x$, differ by the Burgers vector, $\vb$. 
We can consider a case in which $[[\vu(x)]]$ is zero at $x=-\infinity$ and is
$\vb$ at $x=+\infinity$. The region over which the displacement jump
varies substantially defines the dislocation core. 
However, if Shockley partials are formed
$[[\vu(x)]]$ in the dislocation core will not be parallel to $\vb$, as the 
Burgers vectors of the Shockley
partials contain equal and opposite components perpendicular to $\vb$.
We therefore write $[[\vu(x)]] = [[\vu_{\parallel}(x)]] 
                     +[[\vu_{\perp}{x}]]$, the parts of $[[\vu]]$ 
parallel and perpendicular to $\vb$, respectively. 

Using the extrapolation scheme described above, we have estimated
the point at which the parallel portion of the displacement jump 
across the slip plane is $\vb/2$, and take this as the position of 
the dislocation to be used in conjunction with our boundary force 
formula. Our centering of the 
symmetric dislocations (edge and screw), as described above, 
leads to a measured position for the original relaxed
dislocation that is close to zero. (The largest
measured value is 0.02 \angst.) However, the situation for
the mixed dislocations is more complex. The two partials have
different characters, and different widths. Thus our
measured position of the dislocation, which is based on $\vb/2$ is
different than, for example, measuring the positions of $\vb/4$ and $3\vb/4$
and averaging. The position used in estimating the boundary
force for the mixed dislocations is the measured position of $\vb/2$
for the actual configuration, minus the measured position of $\vb/2$ for
the original relaxed dislocation in the unstrained cell.

\subsubsection{\em Measurement of Peierls Stress}
The force per unit length of dislocation needed to
move a dislocation out of the Peierls well (and hence
continuously) in the infinite crystal is estimated as follows.
Starting from (roughly) the center of a well the stress is
increased until the dislocation makes a jump in position which is consistent
with moving to the next well. The lattice resistance
measured at the step before this jump is taken as the maximum
lattice resistance. 
This is therefore intended to be a lower
bound, in the sense that the exact point of the jump might have
been at any applied stress between the one used for the step 
before the jump and the applied stress used for the step during which
the jump occurred.
This estimate of the maximum lattice resistance is computed
using \refeq{force_balance}, including the boundary force term, and
will be called the corrected-Peierls force.
We wish to compare this to the estimate of the 
Peierls force we would have made if we had failed to correct for
boundary forces. The total applied force, for the step just prior
to the jump, is therefore taken as an estimate of the
uncorrected-Peierls force. [Where by corrected and uncorrected we
indicate whether or not the boundary force term in \refeq{force_balance}
is used.]

\subsection{Results and Discussion}
Figures \ref{fig:screw_app}, \ref{fig:m211_app}, \ref{fig:m101_app},
and \ref{fig:edge_elas_app} illustrate the importance of the boundary
force corrections in our simulations by comparing 
(a) the `uncorrected lattice resistance force'
with (b) the corrected lattice resistance force.
The corrections are computed using the coefficients
$A$, $B$ and $C$ in \refeq{PowersEquation} calculated in anisotropic
linear elasticity using the finite element method. The Shockley partials
were assumed to be separated by the distance consistent with 
elasticity theory \cite{teutonico63} in computing the coefficients.
The corrections are the most critical in the case of the
edge dislocation, \reffig{edge_elas_app}, and least critical for the
screw dislocation, \reffig{screw_app}.
Figures \ref{fig:screw_app}a, \ref{fig:m211_app}a, \ref{fig:m101_app}a,
and \ref{fig:edge_elas_app}a plot the position of the dislocation
vs applied force for the four types of dislocation.
For each
dislocation, the maximum applied force simulated was the same for each
size of simulation cell, however to provide more legible graphs the
data where the dislocation has suffered an
excursion larger than 10 \angst away from the
center of the disk are omitted.
The maximum applied force for the 
different dislocations was similar, except for the 60\degree\
dislocation, where a somewhat larger maximum applied force was used.
Figures \ref{fig:screw_app}b, \ref{fig:m211_app}b, \ref{fig:m101_app}b, 
and \ref{fig:edge_elas_app}b show the (negative) sum of the
applied force and the computed boundary force for the same data.
This is therefore the lattice resistance force.
Figures \ref{fig:screw_app}a and \ref{fig:screw_app}b,
for example, are at the same scale.
Notice that for the screw dislocation, the corrections are similar
in magnitude to the lattice resistance being measured,
and so the affect of the corrections is only moderate.
[Notice also in \reffig{screw_app} that for the larger cell sizes the
dislocation, when jumping between wells, does not necessarily land in
the next Peierls well, but may skip over one or more wells. This is
also the case for the other dislocations, but is easiest to see in
\reffig{screw_app}.] 
For the
edge dislocation, the boundary force corrections are much larger
than the lattice resistance, but still provide a reasonably
uniform plot across 13 wells in the case of the 90 \angst
radius cell. The elasticity calculation of A for the edge
dislocation gives values of A that are 3\% to 5\% larger than are optimal
in doing the corrections.
This is visible as the difference between Figures
\ref{fig:edge_elas_app}b and \ref{fig:edge_fudge_app}b, where \ref{fig:edge_fudge_app}b
results from using an $A$ coefficient for each cell that is fit 
rather than determined from linear elasticity.
The key point to be taken away from this series of plots is the recognition
that in the absence of any correction for the effects of boundary force, 
the data gives an impression of a spurious position dependence of the 
lattice resistance curve. On the other hand, once the boundary force 
contribution is removed,
we see that the lattice resistance profiles are nearly identical from one
well to the next {\it and} exhibit a relative insensitivity to the size of
the computational cell. 

\begin{figure}
\epsfxsize=\hsize \epsfbox{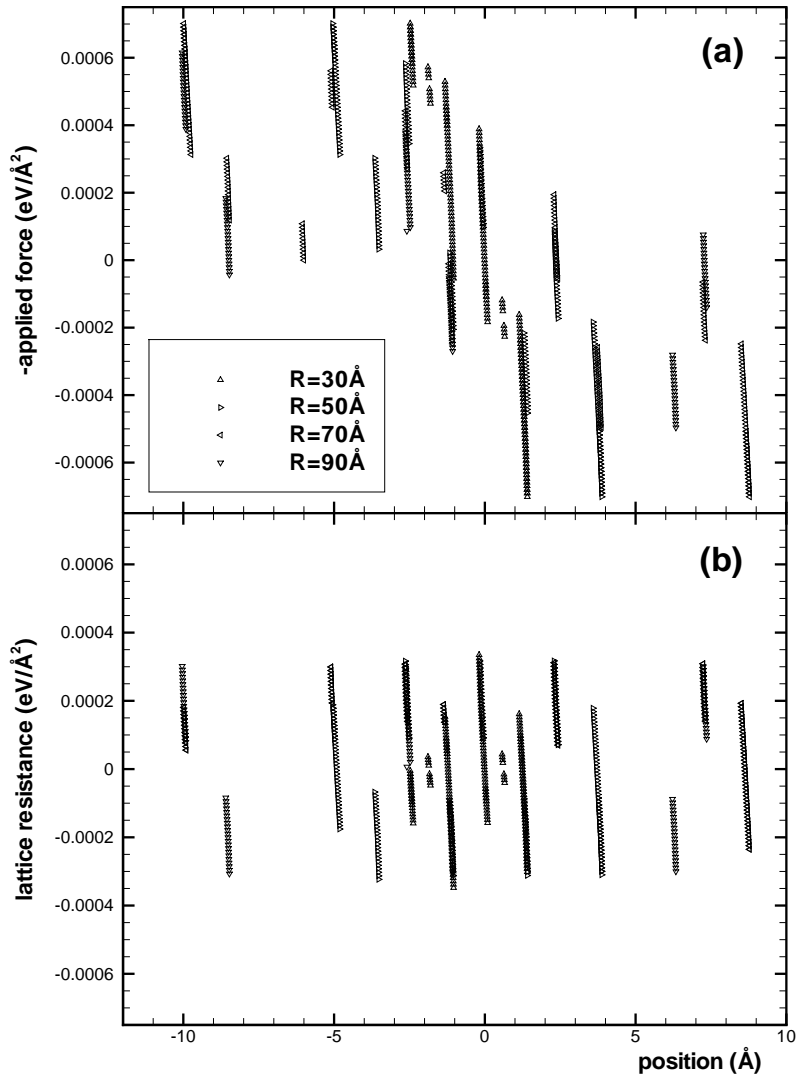}
\caption{\label{fig:screw_app}
Comparison of (a) the applied force and (b) the corrected lattice resistance
for the screw dislocation. 
The $x$-axis shows the position of the dislocation corresponding to where
the measured displacement jump is $b/2$.
}   
\end{figure}
\begin{figure}
\epsfxsize=\hsize \epsfbox{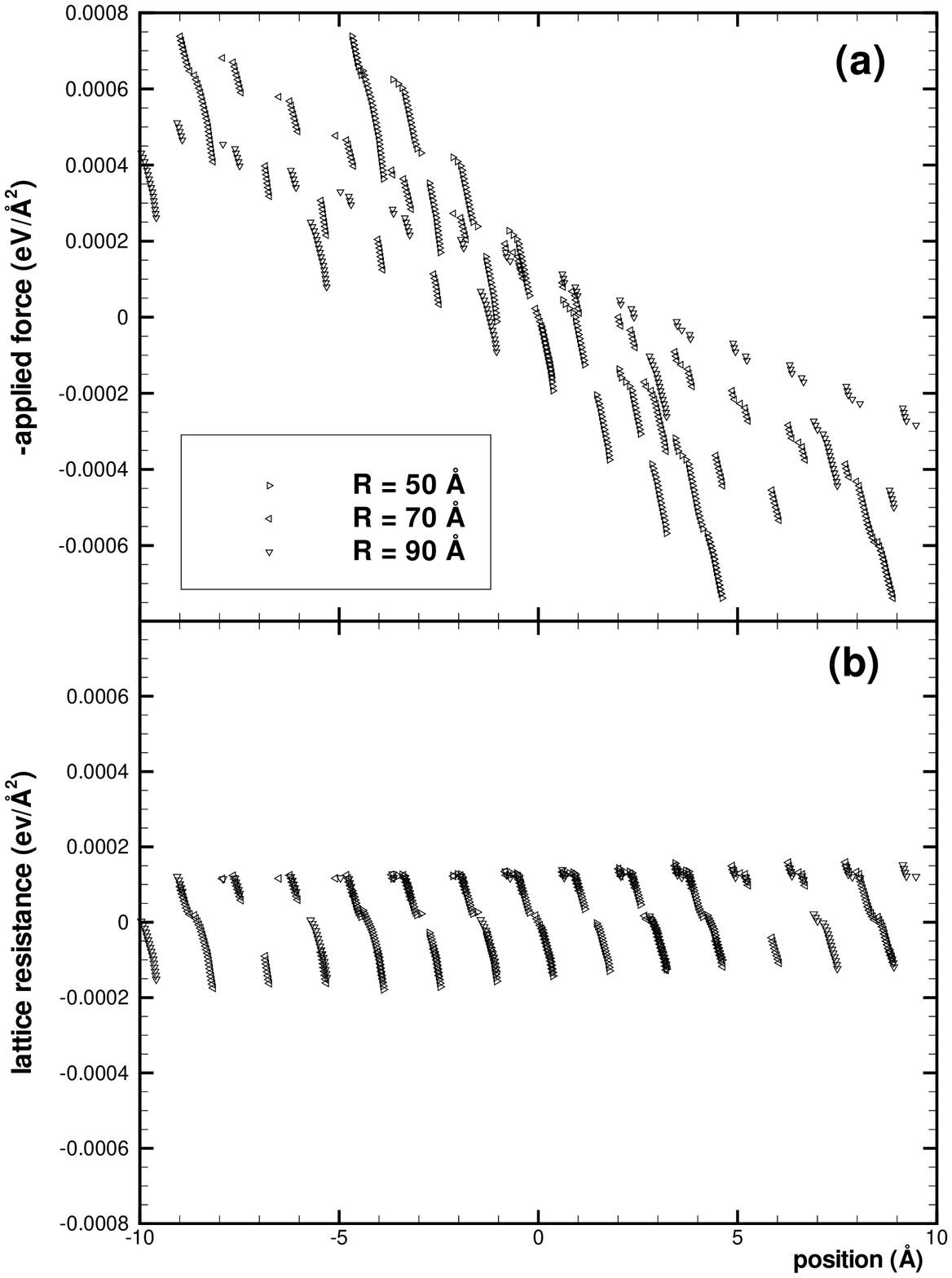}
\caption{\label{fig:m211_app}
Comparison of (a) the applied force and 
(b) the corrected lattice resistance 
for the 30$\degree$ dislocation. The $x$-axis shows the position of
the dislocation corresponding to
where the measured displacement jump is $b/2$,
adjusted so that the position of the dislocation in the
absence of an applied stress is zero.
}   
\end{figure}
\begin{figure}
\epsfxsize=\hsize \epsfbox{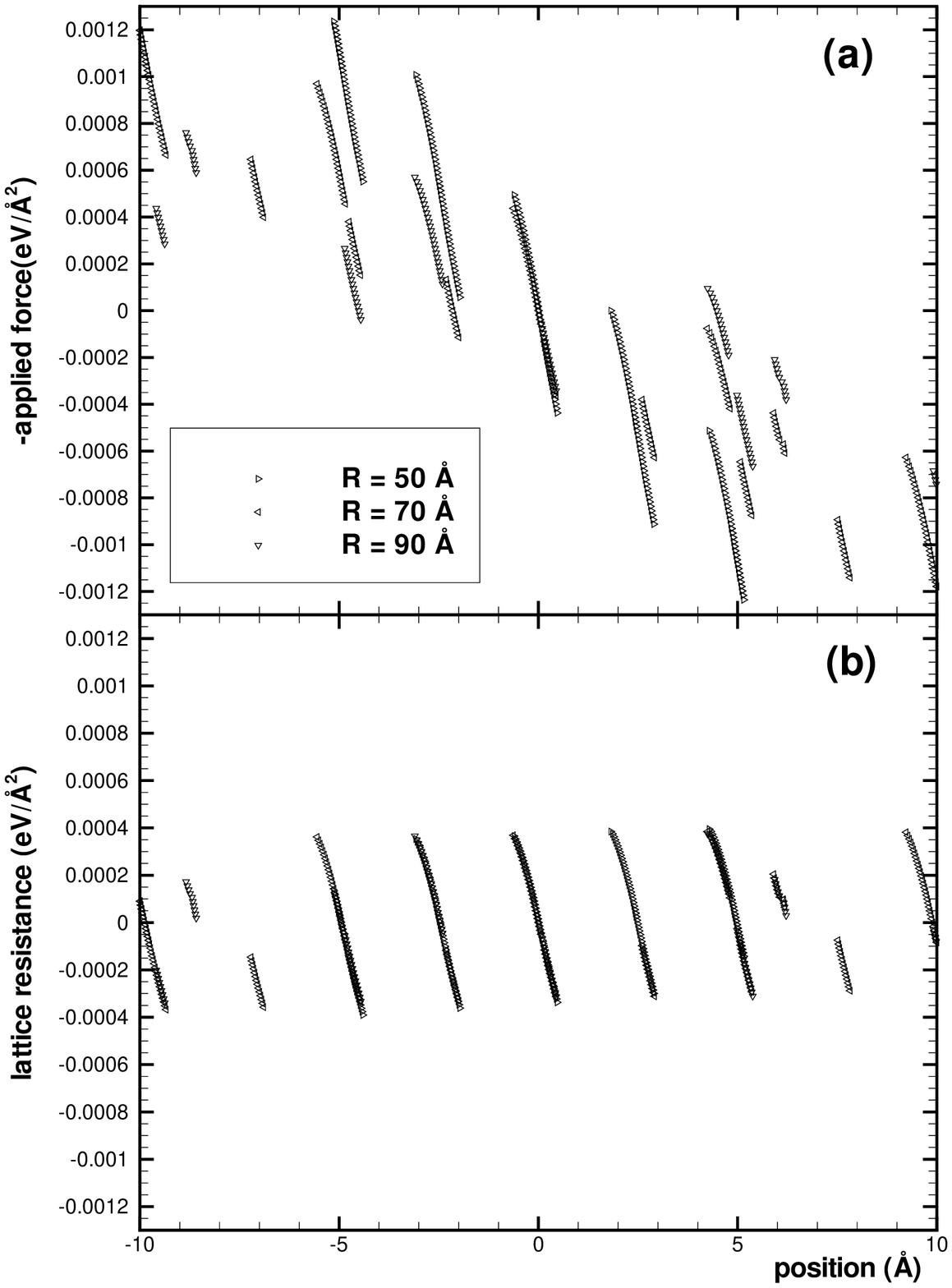}
\caption{\label{fig:m101_app}
Comparison of (a) the applied force and 
(b) the corrected lattice resistance 
for the 60$\degree$ dislocation. The $x$-axis shows the position of
the dislocation corresponding to
where the measured displacement jump is $b/2$,
adjusted so that the position of the dislocation in the absence of
an applied stress is zero.
}   
\end{figure}
\begin{figure}
\epsfxsize=\hsize \epsfbox{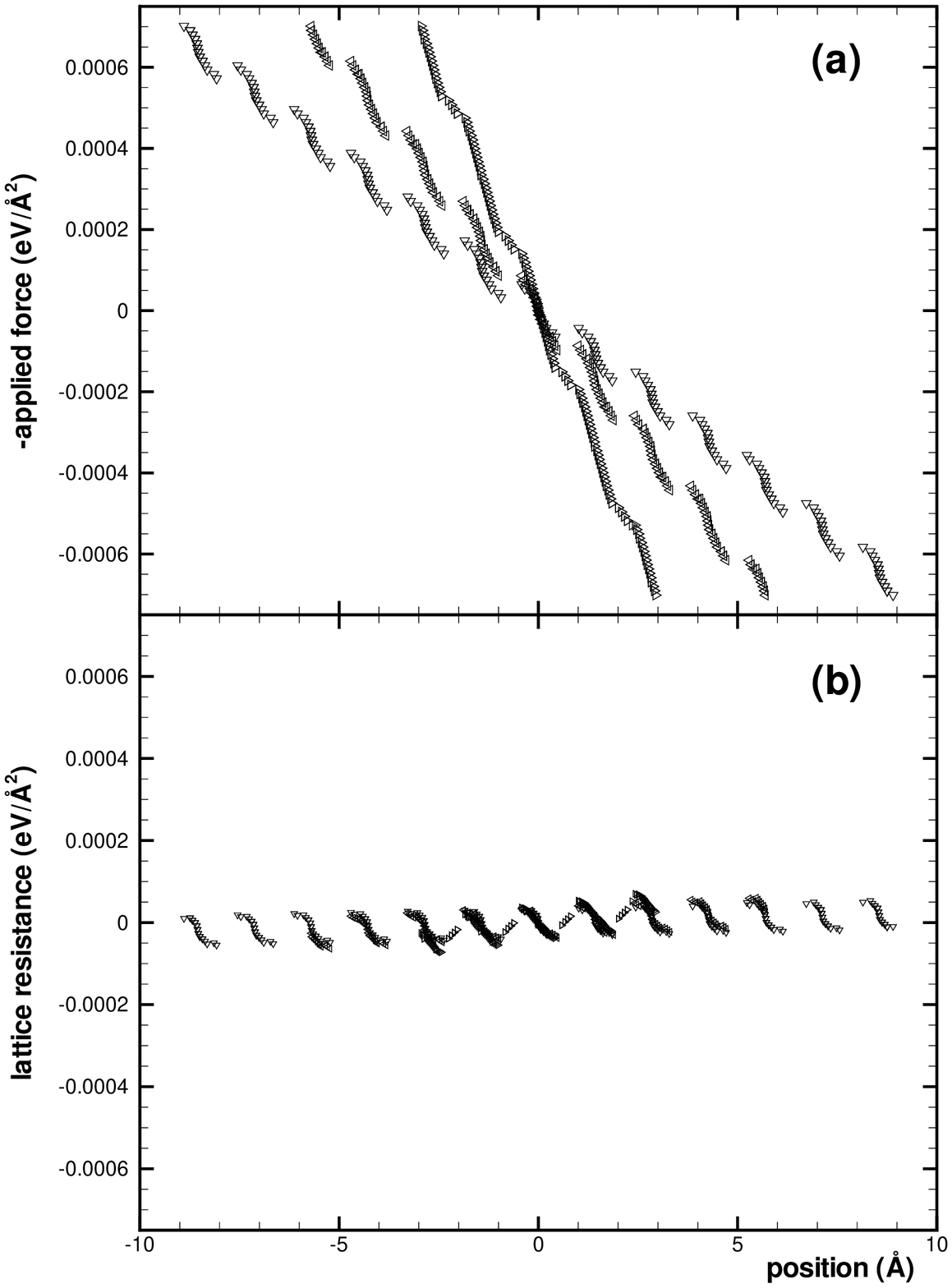}
\caption{\label{fig:edge_elas_app}
Comparison of (a) the applied force and 
(b) the corrected lattice resistance 
for the edge dislocation. The $x$-axis shows the position of
the dislocation corresponding to
 where the measured displacement jump is $b/2$.
}   
\end{figure}
\begin{figure}
\vspace*{-0.5in}
\epsfxsize=\hsize \epsfbox{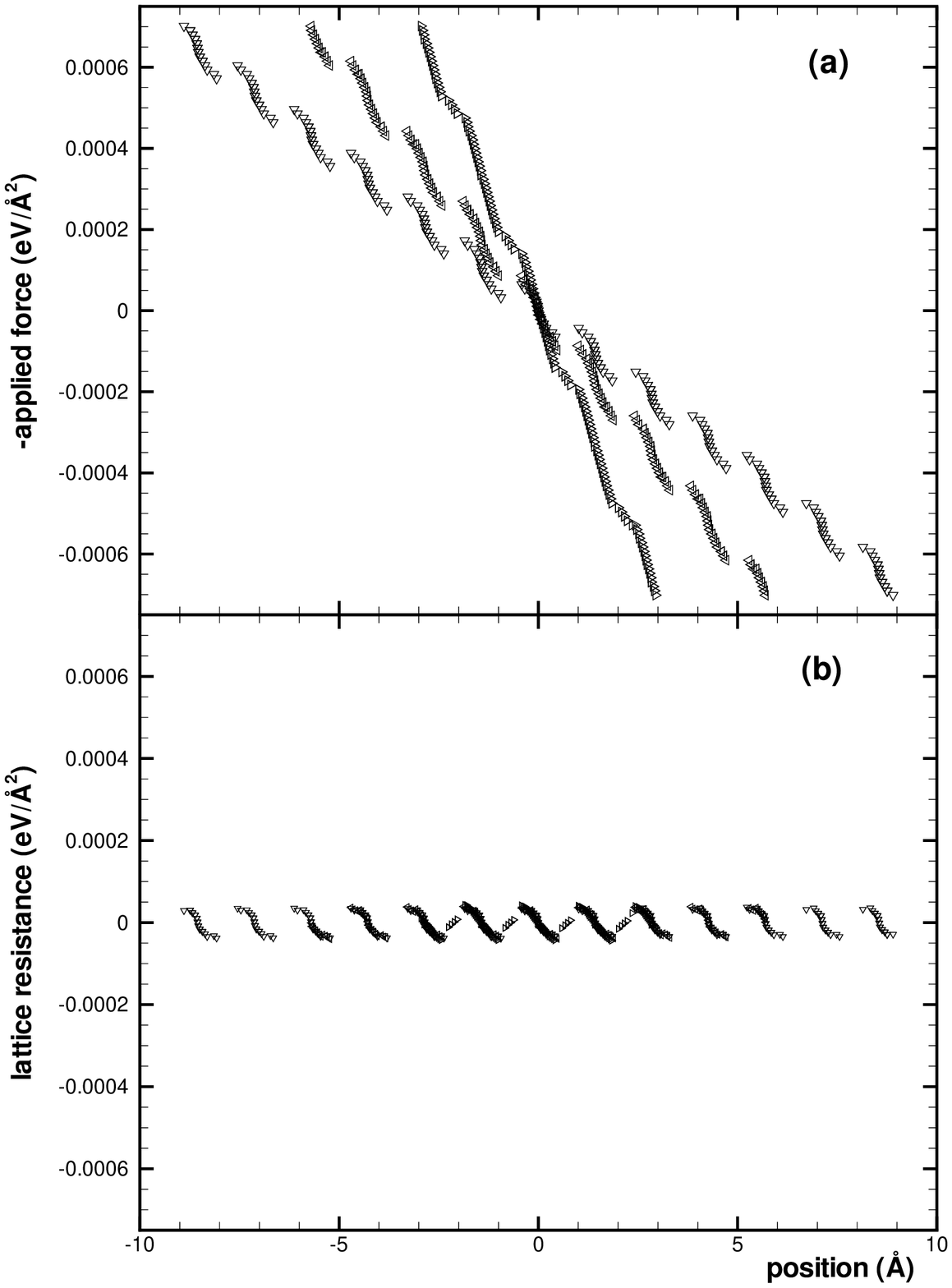}
\caption{\label{fig:edge_fudge_app}
Comparison of (a) the applied force and 
(b) the corrected lattice resistance
for the edge dislocation, where the corrected lattice
resistance is based on optimal choice of the
boundary force coefficient. The $x$-axis shows the position of
the dislocation corresponding to
where the measured displacement jump is $b/2$.
 The adjustment in the boundary force coefficient varies from a
5\% reduction in $A$ for $R=50$ \angst to a
3\% reduction for $R=90$ \angst.
}   
\end{figure}

One of the assumptions underlying our approach is that the lattice resistance
term in \refeq{force_balance} is periodic in $d$ once the boundary
force correction is made. By examining the degree to which our corrected
lattice resistance force is indeed periodic, we test the accuracy of our 
boundary force corrections. (In combination with the assumptions embodied in 
\refeq{force_balance}, since their failure would also introduce errors.)
We can test this visually by graphing the data from one well, offset by the
repeat distance, on top of another well.
Figures \ref{fig:screw_well}, \ref{fig:m211_well},
\ref{fig:m101_well}, \ref{fig:edge_fudge_well}, and \ref{fig:edge_elas_well}, 
show the lattice resistance curves formed by plotting all of 
the wells on top of each other, offsetting each well by an appropriate
number of repeat distances. Except for the edge dislocation, the
boundary force corrections are computed with the coefficients
determined using linear elasticity.
\reffig{edge_elas_well} shows the results for the edge dislocation
with the coefficients computed from elasticity. The slight overestimate
of $A$ is quite apparent in the scatter. \reffig{edge_fudge_well}
shows the same simulation data, but with the boundary-force corrections
based on $A$ values for each $R$ chosen to produce the least scatter.
The values chosen range from 95\% ($R=50$ \angst) to 97\% ($R=90$ \angst)
of the elasticity values. 
\begin{figure}
\epsfxsize=\hsize \epsfbox{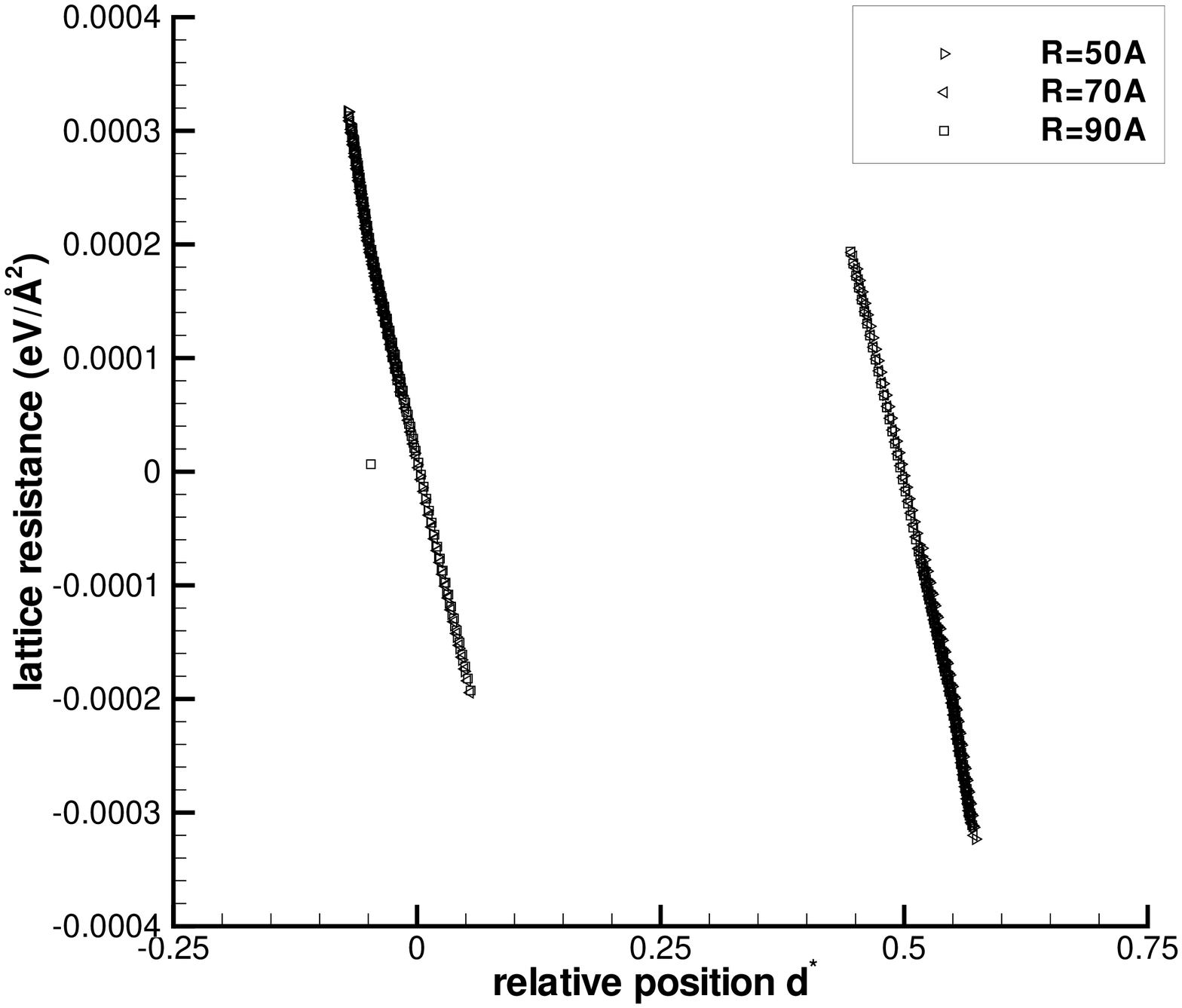}
\caption{\label{fig:screw_well}
Lattice resistance for the screw dislocation. 
Position is plotted assuming the expected periodicity of 
$d_0 = 2.469$ \angst. The data is plotted so that data
expected to be equal by periodicity lies at the same abscissa.
$d^*$ is $d/d_0$ offset by the
integer such that $-0.25 < d^* < 0.75$.
}   
\end{figure}
\begin{figure}
\epsfxsize=\hsize \epsfbox{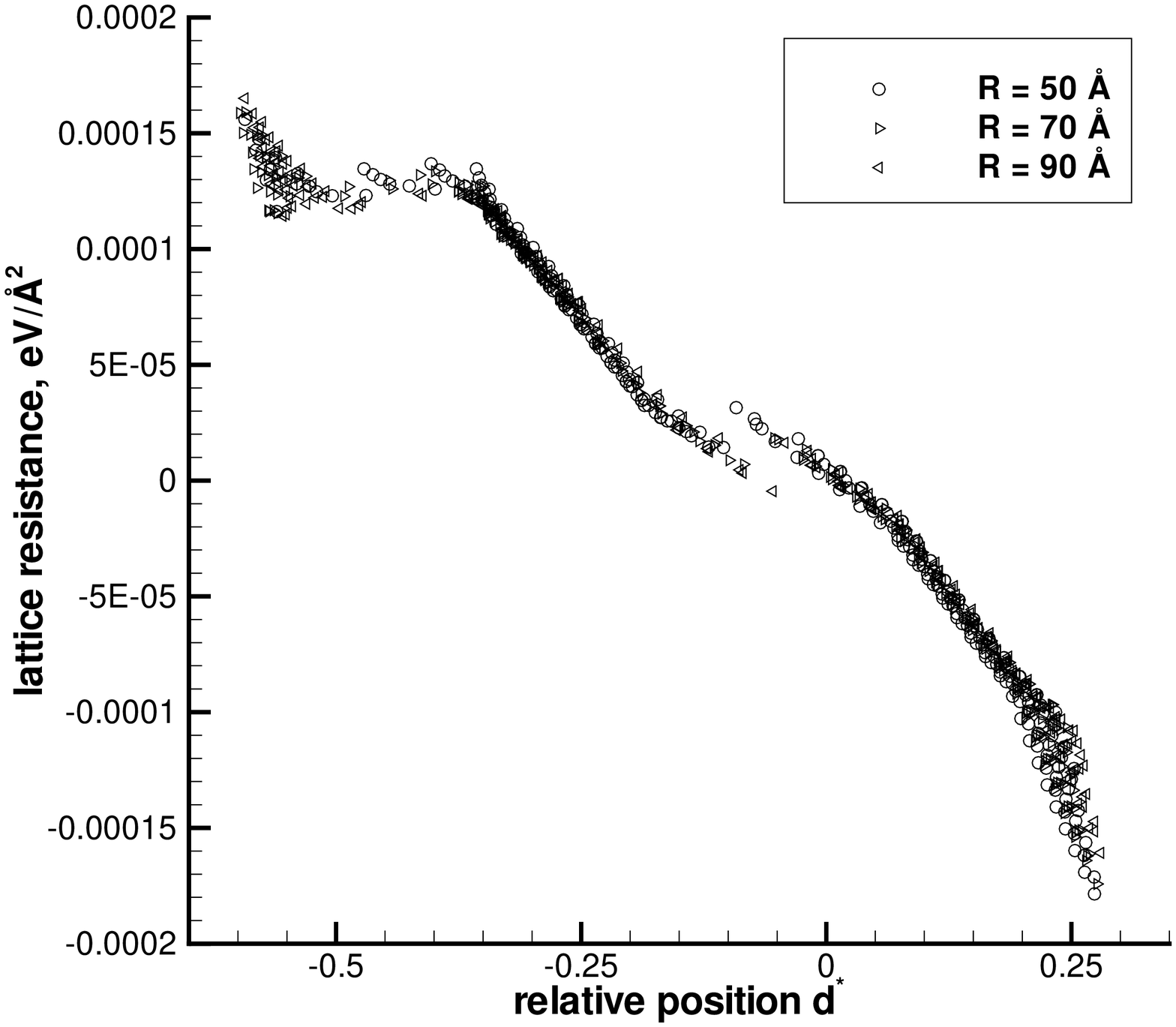}
\caption{\label{fig:m211_well}
Lattice resistance for the 30$\degree$ dislocation. 
Position is plotted assuming the expected periodicity of
$d_0 = 1.426$ \angst. The data is plotted so that data
expected to be equal by periodicity lies at the same abscissa.
$d^*$ is $d/d_0$ offset by the
integer such that $-0.65 < d^* < 0.35$.
}   
\end{figure}
\begin{figure}
\epsfxsize=\hsize \epsfbox{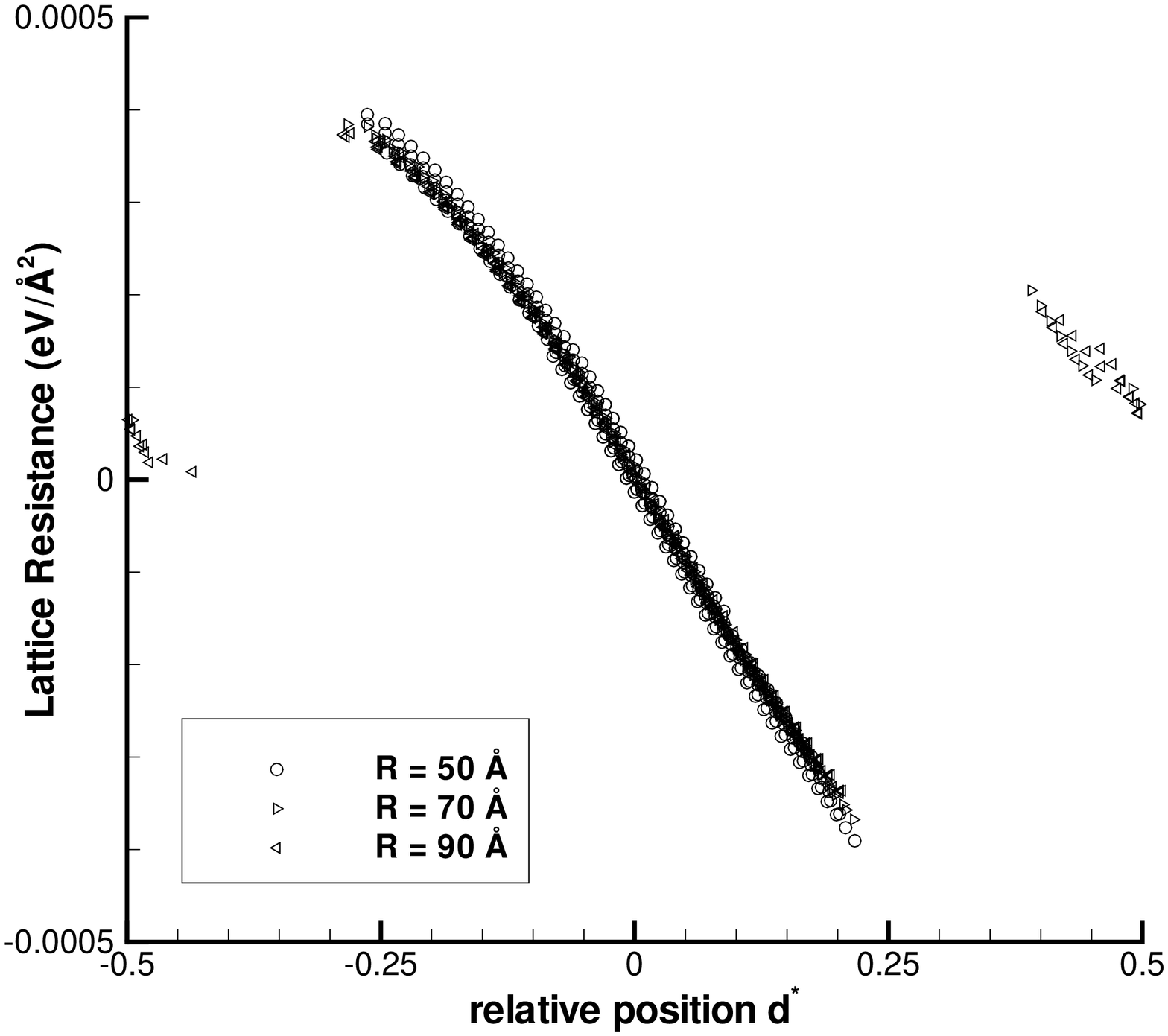}
\caption{\label{fig:m101_well}
Lattice resistance for the 60$\degree$ dislocation. 
Position is plotted assuming the expected periodicity of
$d_0 = 2.469$ \angst. The data is plotted so that data
expected to be equal by periodicity lies at the same abscissa.
$d^*$ is $d/d_0$ offset by the
integer such that $-0.5 < d^* < 0.5$.
}   
\end{figure}
\begin{figure}
\epsfxsize=\hsize \epsfbox{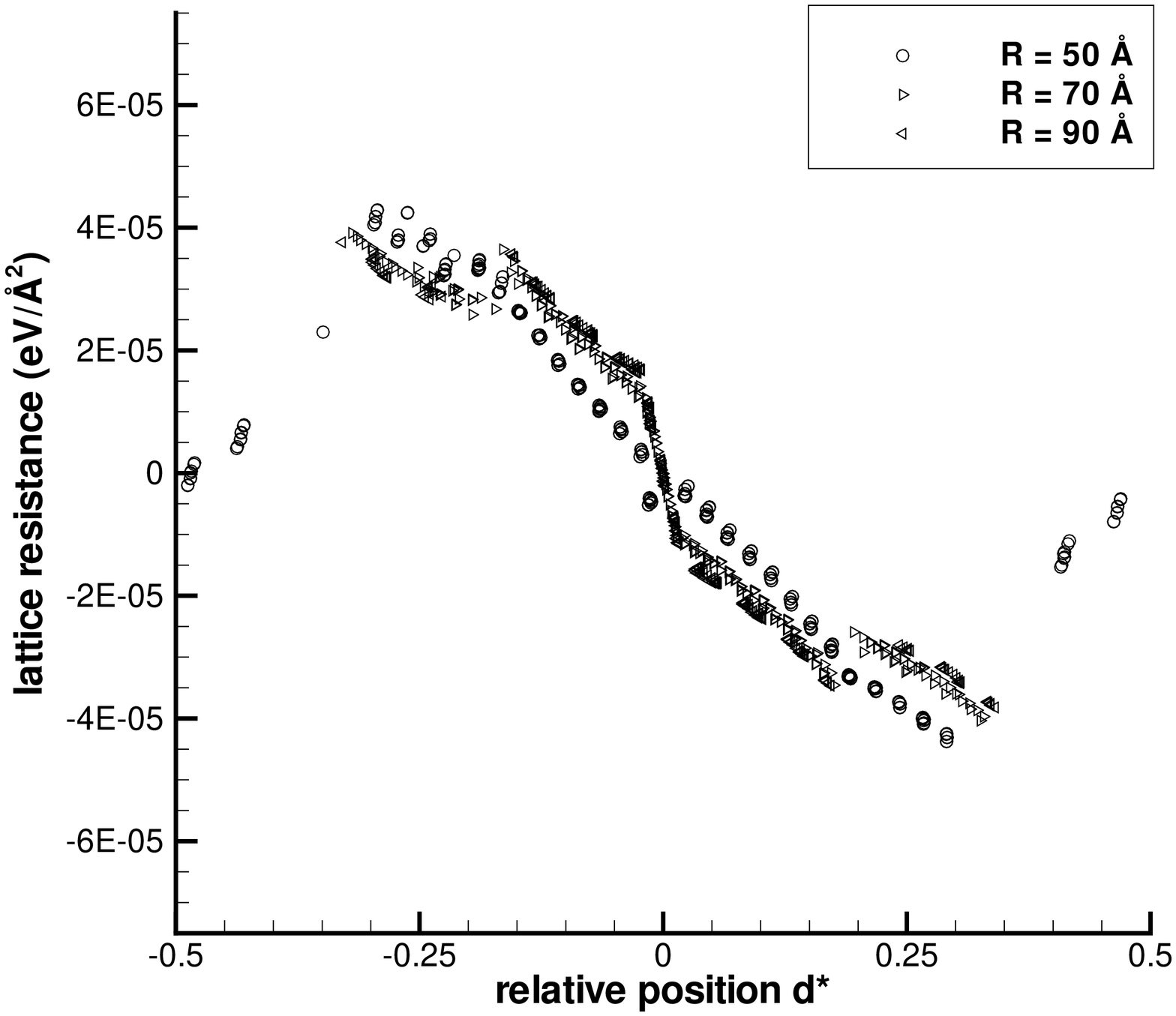}
\caption{\label{fig:edge_fudge_well}
Lattice resistance for the edge dislocation. 
The $A$ coefficients have been adjusted for best fit.
Position is plotted assuming the expected periodicity of
$d_0 = 1.426$ \angst. The data is plotted so that data
expected to be equal by periodicity lies at the same abscissa.
$d^*$ is $d/d_0$ offset by the
integer such that $-0.5 < d^* < 0.5$.
}   
\end{figure}
\begin{figure}
\epsfxsize=\hsize \epsfbox{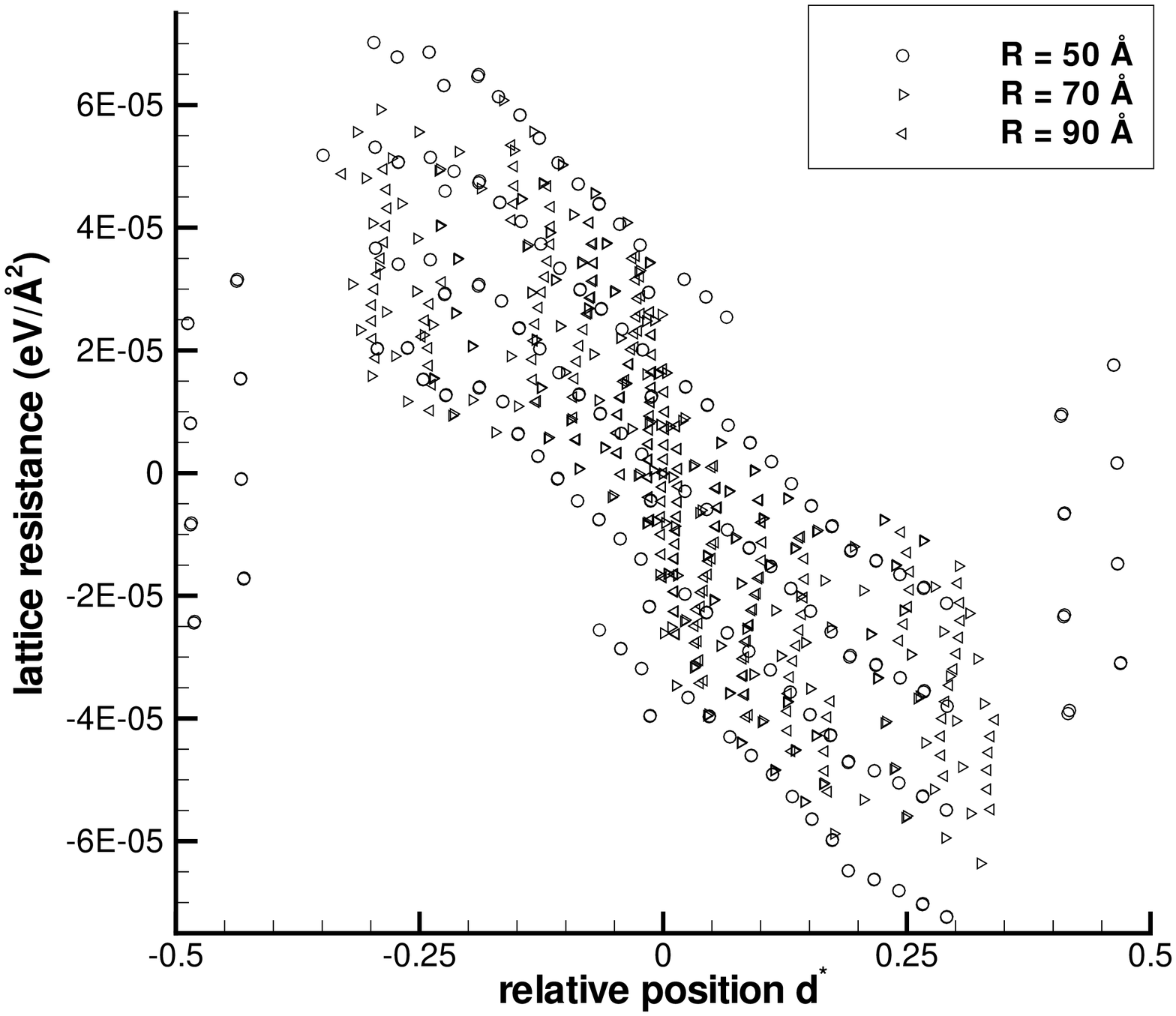}
\caption{\label{fig:edge_elas_well}
Lattice resistance for the edge dislocation.
The $A$ coefficients are from elasticity. 
Position is plotted assuming the expected periodicity of
$d_0 = 1.426$ \angst. The data is plotted so that data
expected to be equal by periodicity lies at the same abscissa.
$d^*$ is $d/d_0$ offset by the
integer such that $-0.5 < d^* < 0.5$.
}   
\end{figure}
\begin{figure}
\epsfxsize=\hsize \epsfbox{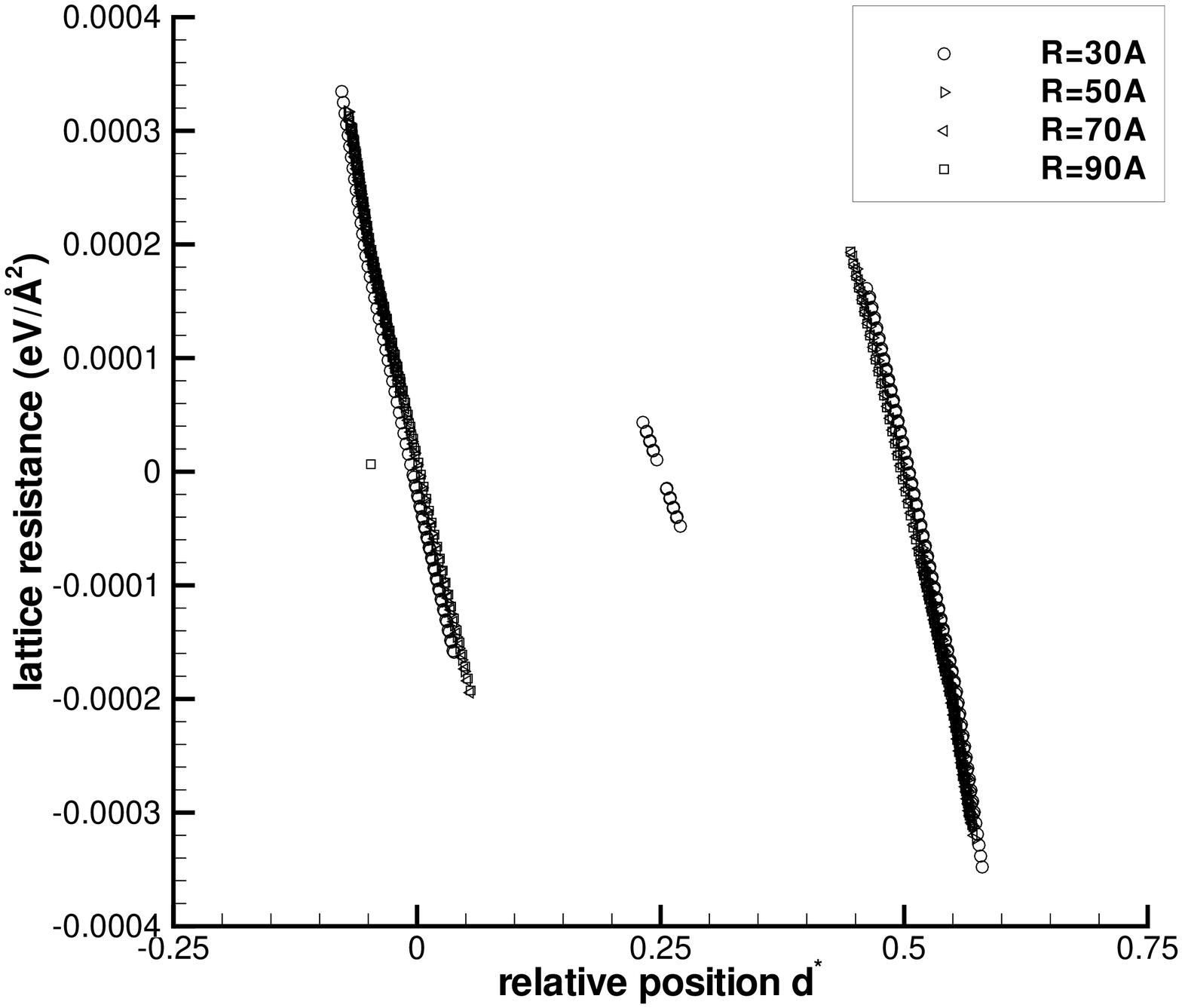}
\caption{\label{fig:screw_well_w30}
Lattice resistance for the screw dislocation, including additional
smaller cell size. 
Position is plotted assuming the expected periodicity of 
$d_0 = 2.469$ \angst. The data is plotted so that data
expected to be equal by periodicity lies at the same abscissa.
$d^*$ is $d/d_0$ offset by the
integer such that $-0.25 < d^* < 0.75$.
}   
\end{figure}

The lattice resistance curves for the different dislocations
are very different, both in shape and in scale.
The screw and edge curves show a reflection
symmetry that the mixed dislocations do not. This is expected,
because the two partials of the edge (screw) dislocation
have the same character.
In all four plots there is good agreement between different
wells for the same radius simulation cell. Except for the edge
dislocation, there is also very good agreement between the
different size simulation cells. A more significant size effect is apparent
in the shape of the curve for the edge dislocation. Because the
lattice resistance forces for the edge dislocation have the smallest
magnitudes, while the boundary forces being subtracted from the
applied force are the largest, some of this discrepancy 
may simply arise from
the fact 
that the radii used for the simulation cells are effectively 
smaller for the edge dislocation. 
In an attempt to shed light on this possibility
a screw dislocation run was made in a cell with a radius of 30 \angst.
\reffig{screw_well_w30} shows this data along with the other 3 radii.
While there is poorer agreement with the other sizes for the 
30 \angst radius, the shape is unchanged. This suggests that 
the size dependence of the shape of the curve for the edge
dislocation may truly differ from that of the other dislocation
studied.\footnote{
The additional points between the two wells for the 
30 \angst screw dislocation do not appear for the
larger cell sizes. Because of the strong boundary forces at
this small size the boundary force can 
be of the same order of magnitude as the 
lattice resistance force in \refeq{force_balance}, and so there
can be solutions in portions of the lattice resistance landscape that
would normally be inaccessible.
For the edge and 60\degree\ dislocations, where the lattice
resistance forces are lower, similar points can be seen for the
50 \angst cell in the case of the edge, and in the larger cell
sizes for the 60\degree\ dislocation. The single ``outlier''
point for the screw dislocation in the 90 \angst cell is
from a simulation that failed to completely converge during the
maximum number of steps allowed. 
}

Perhaps the plot for the 60$\degree$ dislocation looks most
reminiscent of the accessible portion of a sinusoidal
lattice resistance curve. The 30$\degree$ dislocations
shows considerably more structure. The lattice resistance
curve for the screw dislocation, shows two interesting phenomena.
First, the form of the force curve implies that the Peierls
well depicted is split into two separate sub-wells, separated by a local
maximum. Secondly, as the lattice resistance force approaches its
maximum there is no softening of the slope. While we admit to finding
this latter circumstance somewhat perplexing, 
it should be borne in mind that the
force being plotted is the configurational force on the dislocation,
not the actual force on any atom, nor the actual elastic stress at
any point. 

Our primary purpose here in measuring the lattice resistance
curves for the different dislocations is to demonstrate the
ability of the boundary force correction scheme proposed in (I) to
handle edge and mixed dislocations. As we have measured these
curves using a single semi-empirical classical potential, it
is perhaps overly optimistic to present these results
 as a trustworthy guide, even
in qualitative terms, to the variation of lattice resistance
curves with character of dislocation in real aluminum.	 On
the other hand, our results provide substantial quantitative
support for the {\it technique} being used to obtain such curves
and suggests that this same strategy could be useful in the context of
more reliable descriptions of the total energy.

\subsubsection{\em Peierls Stress}
We have measured the Peierls stress, that is the resolved shear stress
in the glide direction required to move a dislocation, both
with and without boundary corrections.
Figures
\ref{fig:screw_peierls}, \ref{fig:m211_peierls}, \ref{fig:m101_peierls} 
and 
\ref{fig:edge_peierls} show the uncorrected and the corrected 
estimates of the Peierls stress as functions of $1/R^2$, for the four
dislocaions studied.
In these plots we show the resolved applied stress 
for the simulation step just {\em before} the jump as the data point.
As an indication of the range that the measurement
might fall in because of step size we show, as an error bar,
the applied stress step size for the uncorrected-Peierls stress, and
an estimate of the effective step size for the corrected-Peierls stress.
In each case, the corrected measurements are substantially less
affected by the size of simulation cell chosen than the 
uncorrected measurements. For the
screw dislocation, where the Peierls force is relatively large,
and the boundary force correction relatively small, 
the uncorrected results at the larger simulation sizes 
are similar to the corrected ones. 
On the other hand, for the edge dislocation, where the boundary
forces are larger, and the Peierls force much smaller, the
uncorrected values are very different from the 
corrected ones.

For the lattice resistance curves above, taking data from wells
many repeat distances away from the center of the cell, obtaining the
agreement shown required care in determining the quadratic
term of the boundary force correction, including tuning
for the edge dislocation. For measuring the Peierls
force using the central well, the boundary force correction is
important, but
the results are not
very sensitive to changes in $A$ of a few percent.

\reffig{peierls} shows our results 
for the Peierls stress in aluminum,
as predicted by the Ercolessi and Adams glue potential. We have
not made any tests of the sensitivity of these results to the
type of potential used, or its details. Nor are we aware of any
experimental data or theoretical predictions for the relationship 
between the 
Peierls stresses for the edge and screw dislocations
in aluminum. Nonetheless,
we find it interesting to see a difference as great as a factor of
8 between the screw and edge values in an fcc metal.

\begin{figure}
\epsfxsize=\hsize \epsfbox{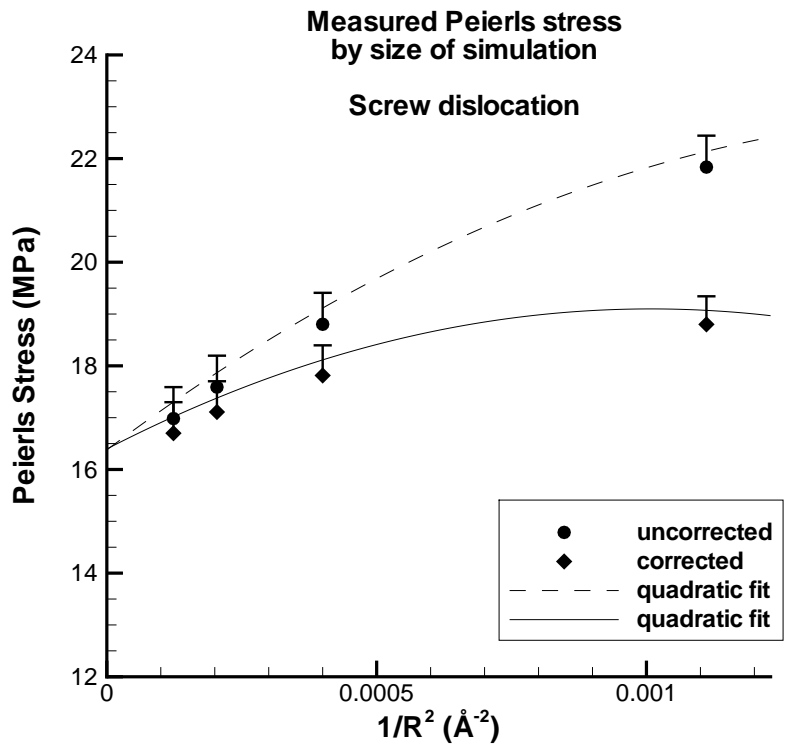}
\caption{\label{fig:screw_peierls}
Peierls stress measured with and without boundary force 
corrections for the screw dislocation. Error bars
represent the approximate step size. The fitted curves
are suggestive only, and are not proposed as extrapolations
to large $R$.
}   
\end{figure}
\begin{figure}
\epsfxsize=\hsize \epsfbox{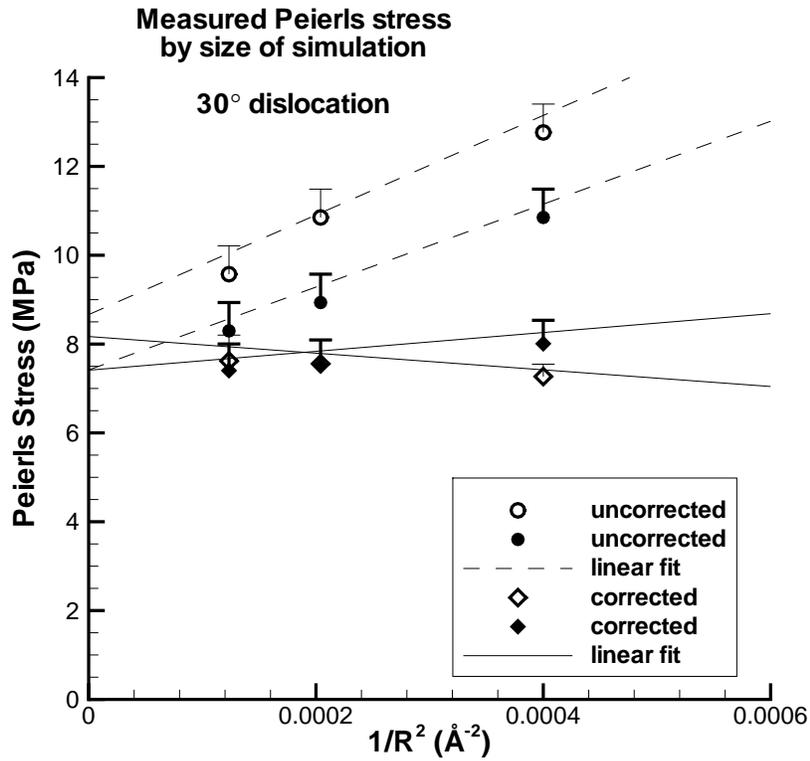}
\caption{\label{fig:m211_peierls}
Peierls stress measured with and without boundary force 
corrections for the 30$\degree$ dislocation. 
Open and solid symbols are the two directions of motion.
The structure
near the maximum on the left side of figure \protect\ref{fig:m211_well}
causes some inconsistency between the $R=50$ \angst corrected value
in that direction and the larger cells.
The fitted lines are suggestive only, and are 
not proposed as extrapolations to large $R$.
}   
\end{figure}
\begin{figure}
\epsfxsize=\hsize \epsfbox{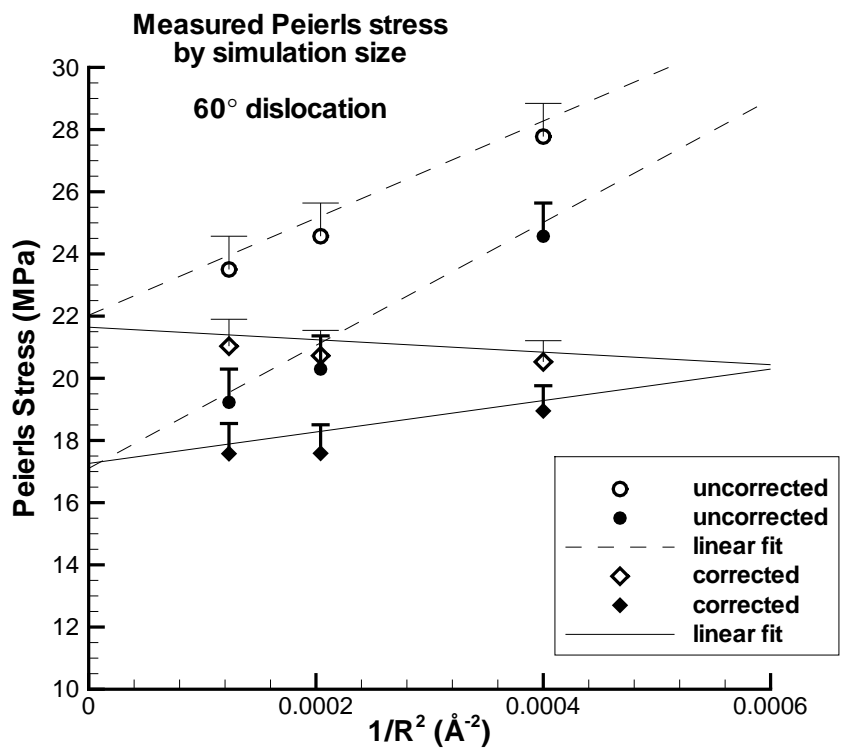}
\caption{\label{fig:m101_peierls}
Peierls stress measured with and without boundary force 
corrections for the 60$\degree$ dislocation.
Open and solid symbols are the two directions of motion.
The fitted lines are suggestive only, and are 
not proposed as extrapolations to large $R$.
}   
\end{figure}
\begin{figure}
\epsfxsize=\hsize \epsfbox{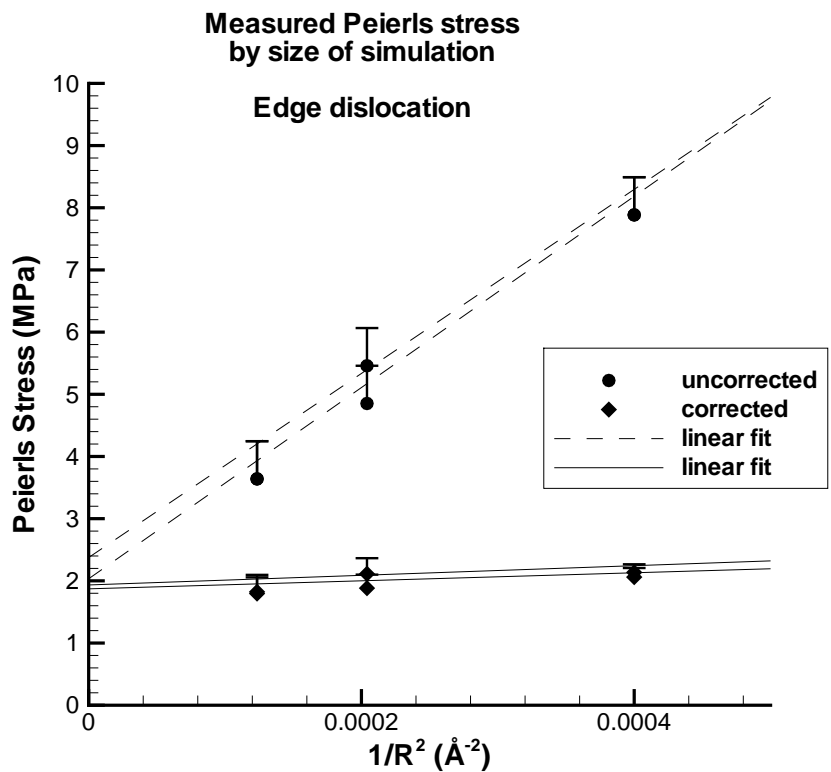}
\caption{\label{fig:edge_peierls}
Peierls stress measured with and without boundary force 
corrections for the edge dislocation.
Data for the two directions of motion are shown,
but should be equivalent for the edge dislocation.
The fitted lines are suggestive only, and are 
not proposed as extrapolations to large $R$.
}   
\end{figure}
\begin{figure}
\epsfxsize=\hsize \epsfbox{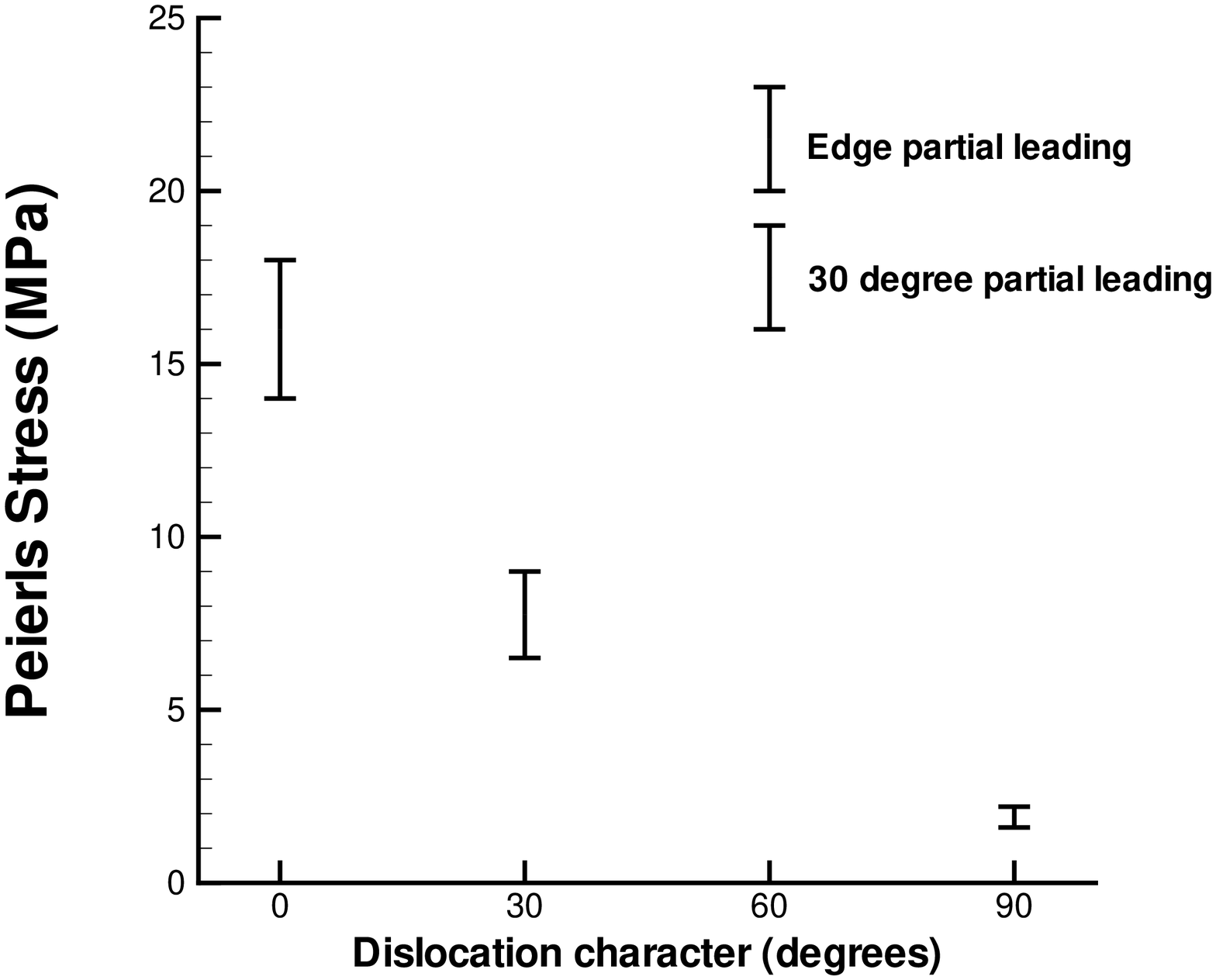}
\caption{\label{fig:peierls}
Peierls stress for the different dislocations. 
The dislocations are labeled by the angle
between the line direction
and the Burgers vector, with the screw dislocation labeled 0 and the
edge dislocation labeled 90.
The range shown for each measurement was 
subjectively determined by examining Figures
\protect\ref{fig:screw_peierls}, 
\protect\ref{fig:m211_peierls}, 
\protect\ref{fig:m101_peierls} 
and 
\protect\ref{fig:edge_peierls}.
For the 60$\degree$ dislocation the dislocation core is not
symmetric with respect to the direction of motion, because
one partial has edge character, while the other partial has 
mixed character. As it moves from one Peierls well to the next
it will cross an identical maximum energy in either direction, but
as the shape of the hill need not be the same on both sides, the
maximum resistance encountered in the two directions differs.
In principle the two directions of motion for the 30$\degree$
dislocation are different, but the data does not show a clear
difference in Peierls stress.
}   
\end{figure}

J. P. Simmons, et al., measured the Peierls stress of unit
$\half a_0 \{1 1 0\}$ dislocations in an ordered $L1_0$ structure
in several embedded-atom method potentials
which were fitted to
$\gamma$--TiAl \cite{simmons97}. 
For their preferred potential they measured the Peierls
stress for the four dislocation characters treated in this work.
Those Peierls stresses are about a factor of 10 larger than in our
material, but the overall qualitative result that the maximum lattice
resistance for the screw and $60^{\circ}$ dislocations are significantly
larger than those of the edge and $30^{\circ}$ dislocations is
similar to what they observed. They ascribed this overall result
to the difference in the density of atom planes in the direction of
motion, which is a factor of 1.7 higher for the edge and $30^{\circ}$
dislocations. They support this suggestion based on the
Peierls-Nabarro model, using the work of 
Weertman and Weertman \cite{weertman92}. The suggestion that this
is the primary element in the overall differences between the
different dislocation characters is to some degree 
supported by the fact that
we see, in an fcc potential with much smaller Peierls stresses, 
the same type of difference between the two dislocations with
the larger distance between atom planes in the direction of motion
and the two with the smaller distance,
but the relationships between the Peierls stresses of dislocations
with the same distance between atomic planes are not similar between
the two cases. However, in our case the Peierls stress for the
$30^{\circ}$ dislocation exceeds that of the edge dislocation
by almost as much as the
difference between the Peierls stresses for the $30^{\circ}$ dislocation
and the screw dislocation, 
and so the difference in atomic plane density does
dominate the results as clearly as in the Simmons et al. results. 

Our result for the Peierls stress of the screw dislocation is shown as
14-18 MPa, and is 20\% or more smaller than the estimate in (I) for the same
material, using essentially the same approach. To some extent this
involves the inclusion of a larger disk size than in (I).
There is also a significant effect in this case from the use of the
linear elastic solution for
the two Shockley partials in generating the positions of the atoms
in the fixed ring, as opposed to the use in (I) of the linear elastic
solution for the perfect dislocation.

Bulatov, et al., have also measured the Peierls stress of the screw and
$60^{\circ}$ degree dislocations for the same Ercolessi and Adams
potential for Aluminum, using fully periodic boundary 
conditions \cite{bulatov99}. 
They report a Peierls stress of 82 MPa for the screw
dislocation, and 47 MPa for the $60^{\circ}$ degree dislocation. These
results are substantially higher than ours. They also exceed all of our
uncorrected estimates, which should definitely be overestimated because of
the boundary repulsion. We can see no reason why our estimates should be
so substantially underestimated as the Bulatov, et al. figures would suggest.
Our results however, as discussed above, are for a particular stress state, 
chosen to provide equal Peach-Kohler forces on the two partials.

\subsection{Conclusions}
When simulations are performed of dislocations in finite-size
simulation cells, boundary forces will often be important.
One approach to correcting for boundary forces
is that proposed by Shenoy and Phillips in (I). We have
shown how that method can be extended to edge and mixed dislocations,
and to general cell shapes, and have applied the technique
to the measurement of the lattice resistance curves and
Peierls stresses of four mobile dislocations in fcc aluminum.
In addition, we have turned the usual elastic arguments concerning the
nature of configurational forces on their side and introduced a purely
atomistic scheme for computing such boundary forces. A second
outcome of our analysis is the recognition of the consistency of
the boundary forces obtained using either elasticity or atomistic analysis.
We also note that our results are illustrative in that they signify
a large degree of control over the various contributions to the total
energy in a finite size system.

Our methods assume that the total configurational force on a
dislocation can be divided into three parts,
\begin{itemize}
\item{
An applied force given by the Peach-Kohler formula.
}
\item{
A periodic lattice resistance force, independent of position
within the simulation cell, and independent of cell radius.
}
\item{
A boundary-effect force which is a function of $d/R$.
For the case of two Shockley partials the boundary-effect force also depends
on $s/R$, where $s$ is half the partial separation. 
}
\end{itemize}
The data from our simulations support this assumption, except
that for the edge dislocation the shape of the measured
lattice resistance force curve has some dependence on the
size of the simulation cell. In addition we find that
a linear elastic computation of the boundary-force function
gives reasonable results,
which were adequate for our purposes for all of the dislocations
except the edge.
While we have considered the
dependence of $A$ on $s/R$ caused by the extension (treated as
dissociation) of the
dislocation, and have considered the higher order coefficients 
$B$ and $C$ as well, these
are small effects, and we expect that in many cases simply using the
value of $A$ computed for a perfect dislocation will provide
a reasonable estimate of the boundary force.
We also find that both the shape of the lattice resistance landscape,
as well as the magnitude of the Peierls stress, depend strongly on
the dislocation character. 

From the perspective of future applications of ideas like those introduced
here, we note that we foresee at least two fertile avenues for further work.
First, within the context of the type of simple geometries considered here,
we argue that the analysis of boundary forces could be a potentially useful
way of turning the disadvantage of small system sizes used in first-principles
simulations of complex systems into a means of precise control of the
boundary conditions used in such simulations. Perhaps more importantly, 
we view our analysis as a first step towards the more judicious management 
of boundary conditions associated with complex three-dimensional geometries 
involving dislocations. In particular, recent work on both dislocation 
junctions \cite{bulatov98,zhou98,rodney99} 
and cross-slip \cite{rasmussen97, rasmussen97a} 
are a significant first step towards informing higher-level dislocation
dynamics simulations of plasticity on the basis of atomic-level understanding
of the ``unit processes'' involving dislocations. One of the key challenges
faced in effecting calculations that are quantitatively meaningful
is combatting the spurious effects induced by the presence of unwanted nearby
boundaries. The present work offers the possibility of finite-element based,
linear elastic calculations of the effects of such boundaries, thus clearing
the way for substantive quantitative insights into these processes.

\subsection*{Acknowledgments}
We would like to thank Nitin Bhate, Ron Miller, Dnyanesh Pawaskar,
and S.I. Rao for helpful conversations. This work was partially 
supported by the DOE through Caltech's ASCI Center for the Simulation of the
Dynamic Response of Materials; and the NSF under grant CMS-9971922 and 
through the MRSEC at Brown University.

\subsection*{Appendix A --- Partial Splitting}
For our model aluminimum system all of the easy-glide dislocations
split (approximately) into Shockley partials. The width
of the partial dislocations is
similar to their separation distance, so that their core regions
overlap to some extent.
While this means that there is no region on the slip plane
that is a close approximation
to a stacking fault, the elasticity model of two unextended Shockley
partials separated by a stacking fault remains useful.

\subsubsection*{\em Image Solution}
While we are unable to use the image solution to handle the
actual Shockley partials of the screw dislocation, which include edge
components, it is possible to use it for the simpler case of
pure screw partials, each with a Burgers vector of one-half $b$,
at fixed separation. We hope that this simplified model will at 
least indicate the nature of the extra subtlety in the boundary
force as a result of the splitting of the dislocation into
partials.  There are two possible boundary conditions
to consider. One is where the fixed displacements at the boundary
are those of a single screw dislocation at the center of the cell.
In the other case we compute the fixed displacements based on
the two partial dislocations, with the center of the cylinder
located half-way between the two partials.

Letting $s$ be one-half the partial separation, the partials
are located at $d+s$ and $d-s$.
When the boundary conditions are those of a single dislocation,
only two image dislocations are needed. These are screw dislocation at 
$L = \frac{R^2}{d+s}$ and $L = \frac{R^2}{d-s}$, 
each with Burgers vector $b/2$. (When $d=\pm s$, one of the partials
is at the center of the disk, and its image dislocation is
not needed.) The force on the dislocation is now
\be
 F/l = - \frac{\mu (b/2)^2}{2 \pi} \left[
          \frac{1}{ \frac{R^2}{d+s} - (d+s) }
       +  \frac{1}{ \frac{R^2}{d-s} - (d+s) }
       +  \frac{1}{ \frac{R^2}{d+s} - (d-s) }
       +  \frac{1}{ \frac{R^2}{d-s} - (d-s) }  \right].
\end{equation}
While the result of integrating this to obtain the energy is
not particularly complicated, neither is it particularly enlightening.
Letting $c=s/R$, and treating the partial splitting as making the
terms in our expansion of the boundary energy depend on $c$, we
obtain
\bea
   A &= 2 \pi \frac{(1+c^4)}{(1-c^2)^2(1+c^2)}     \\
     &= 2 \pi \left[ 1 + c^2 + 3 c^4 +O(c^6)  \right]     \\
   B &= \pi   \left[ 1 + 4c^2 +O(c^4). \right]
\end{align}
[The result for A under these conditions was shown in
figure A 1 of (I).] 

When the boundary displacements are computed using the partials
(which mimics the boundary conditions of our simulations) the
situation is slightly more complicated. The boundary conditions
are the displacements resulting from one partial at $s$ and one
at $-s$.  The actual dislocations are at $d+s$ and $d-s$.
A $+b/2$ dislocation at $s$ and one at $R^2/s$ combine to give the
boundary displacements of a $+b/2$ dislocation at the center.
A $+b/2$ dislocation at $d+s$ and one at $R^2/(d+s)$ also combine
to give the boundary displacements of a $+b/2$ dislocation at the center.
Thus the actual $+b/2$ partial at $d+s$ together with 
a $+b/2$ dislocation at $R^2/(d+s)$ and a $-b/2$ dislocation at
$R^2/s$ will combine to produce the displacements of the
$b/2$ dislocation at $s$. Handling the other partial in the same way,
we then have four image dislocations, two of like Burgers vector
as the actual partials, and two with equal and opposite Burgers
vectors. There are now eight terms contributing to the configurational
force on the dislocation computed based on the image dislocations, and
we have
\begin{eqnarray}
 F/l  = - \frac{\mu (b/2)^2}{2 \pi}
              \left[
          \frac{1}{ \frac{R^2}{d+s} - (d+s) }           \right.
          & + &  \frac{1}{ \frac{R^2}{d-s} - (d+s) }    
     \notag \\
   \mbox{}    -   \frac{1}{ \frac{R^2}{s}   - (d+s) }  
     &  - & \frac{1}{ \frac{R^2}{-s}  - (d+s) }    \notag \\
   \mbox{}      +  \frac{1}{ \frac{R^2}{d+s} - (d-s) }
     &  + & \frac{1}{ \frac{R^2}{d-s} - (d-s) }    \notag \\
   \mbox{}   -  \frac{1}{ \frac{R^2}{s}   - (d-s) }
    &  - & \left. \frac{1}{ \frac{R^2}{-s}  - (d-s) } \right].
\end{eqnarray}
For these boundary conditions we have
\bea
   A &= 2 \pi \left[ 1 + 3 c^4 +O(c^6) \right]       \\
   B &=   \pi \left[ 1 + 4c^2 +O(c^4)  \right]. 
\end{align}
Aside from the usefulness of this result as a benchmark in testing
our FEM computations, the point of interest is that the highest order
correction to the boundary force as a result of partial splitting, which
is of order $d s^2/R^4$, is zero when the fixed displacements are
computed based on the partials. For the more general case, with 
anisotropic elasticity, and non-screw dislocations 
this term is no longer zero, but it 
is small, and
varies 
in sign between the different dislocations.

Subtracting, rather than adding, the forces on the two partials, and
dividing by two, we obtain the extra force (per unit length) applied
to the ``spring'' connecting the two partials because of the
boundary conditions. This may be of some use in estimating how large
a cylinder is needed to avoid distortions in the partial separation,
assuming the stacking fault energy is available. For the case
where the boundary displacements are based on a perfect dislocation,
there is a force even when the dislocation is at the center of the
disk. The result is that the ``spring'' is compressed by a force 
on each end of
\be
   F_{\subrm{C}}/l = \frac{\mu (b/2)^2}{2 \pi} \frac{2 s^3}{R^4-s^4},
\end{equation}
at the nominal separation of $2s$. This would then be partially
relaxed by a reduction in the separation.
Including the overall displacement of the dislocation, while still keeping 
the partial separation fixed, and keeping only terms to
order $R^{-4}$ the compression is
\be
    F_{\subrm{C}}/l \approx \frac{\mu (b/2)^2}{2 \pi} \left[
         \frac{2 s^3}{R^4} + \frac{2 s d^2}{R^4} \right].
\end{equation}
If the boundary displacements are instead based on the
two ``partials'', the force on each end of the ``spring'' is
\be
    F_{\subrm{C}}/l \approx\frac{\mu (b/2)^2}{2 \pi} \frac{2 s d^2}{R^4},
\end{equation}
to the same approximation. While these results are for the
unphysical case of an isotropic screw dislocation split into
pure screw partials, it should be a correct guide to the
scaling of the dominant term in the expansion.

\subsubsection*{\em Elasticity Solution}
In this section we extend \refeq{Ea} to the case
where the dislocation is split into two (Shockley) partials.
The partials have Burgers vectors adding up to the Burgers
vector of the perfect dislocation
\be
     \vb^{(l)} + \vb^{(r)} = \vb^{(f)},
\end{equation}
where a superscript l(r) will refer to the left(right) partial,
and a superscript f will refer to the full dislocation.
The partials have the same line direction as the perfect dislocation,
and, for the case of Shockley partials, the Burgers vectors
of the partials lie in the same slip plane that the Burgers vector
of the perfect dislocation does.

We apply the sextic formulation of 
Eshelby, et. al \cite{eshelby53,foreman55}, as
applied to the case of partial dislocations by Teutonico 
\cite{teutonico63}.
We follow
the treatment of Hirth and Lothe \cite{sextic} explicitly,
varying the notation just enough to include the superscripts
for the two partials and the perfect dislocation.

The coordinate axes are chosen so that $\hat{x}_3$ is the line
direction and $\hat{x}_2$ is the perpendicular to the slip plane.
This means that $\hat{x}_1$ is the direction in which the
movement of dislocation is measured, and the direction in
which the partial separation is measured. We choose $s$ as
half the distance between the two partials, and assume it to be
independent of $d$. The left and right partials are therefore located at
$d - s$ and $d + s$ respectively. (Notice that this choice of
axes is different from that used elsewhere in this paper.)

We wish to evaluate
\be 
    \dea    = \half \imsp \vt_P \cdot \jup \rd S
          - \half  \ims \vt_O \cdot \juo \rd S.    
                  \label{eq:Ea_formal}
\end{equation}
We write
\bea
 \frac{1}{l} \half \ims \vt_O \cdot \juo \rd S
        &= \half \int_{-R}^{-s -r_c } 
          \left\{ \vt_O^{(l)}(x_1) + \vt_O^{(r)}(x_1) \right\}
             \cdot \vb^{(f)} dx_1        \notag \\ 
       &+ \half \int_{-s+r_c}^{s -r_c } 
          \left\{ \vt_O^{(l)}(x_1) + \vt_O^{(r)}(x_1) \right\} 
               \cdot \vb^{(r)} dx_1    
                                           \label{eq:EAO}    \\
 \frac{1}{l} \half \imsp \vt_P \cdot \jup \rd S
        &= \half \int_{-R}^{d -s -r_c } 
          \left\{    \vt_P^{(l)}(x_1) + \vt_P^{(r)}(x_1) \right\}
                 \cdot \vb^{(f)} dx_1               \notag \\ 
       &+ \half \int_{d-s+r_c}^{d + s -r_c } 
          \left\{    \vt_P^{(l)}(x_1) + \vt_P^{(r)}(x_1) \right\}
                         \cdot \vb^{(r)} dx_1. 
                                                 \label{eq:EAP}
\end{align}
Rewriting equations (\ref{eq:EAO}) and (\ref{eq:EAP}) 
in terms of $\vt^{(l)}(x_l)$ 
where $x_l$ is measured from the location of the left partial, 
and $\vt^{(r)}(x_r)$ where $x_r$ is measured from the location of the
right partial, and taking the difference we have
\begin{equation}
   \dea / l  = 
       \half \int_{-R-d+s}^{-R+s} \vt^{(l)}(x_l) \cdot \vb^{(f)} dx_l
     +  \half \int_{-R-d-s}^{-R-s} \vt^{(r)}(x_r) \cdot \vb^{(f)} dx_r.
                 \label{eq:tau_dot_b}
\end{equation}

The stresses within the sextic formulation are computed as follows
using the notation of \cite{sextic} to which the reader is referred for
the derivations.

Define
\be
   a_{ik} \equiv c_{i1k1} + (c_{i1k2}+c_{i2k1})p + c_{i2k2}p^2,
\end{equation}
for $p$ an arbitrary complex number.
We need solutions of the equations
\be
   a_{ik} A_{k} = 0, \hspace{2em} i=1,2,3,
   \label{eq:define_A}
\end{equation}
which can only hold if $\det([a_{ik}(p)]) = 0$.
This sixth degree real polynomial equation can be shown to have
no real roots, so that its solutions are three pairs of complex
conjugates, and pick $p_{(1)}$, $p_{(2)}$ and $p_{(3)}$ 
to be the three solutions
with positive imaginary parts. We let $\vect{A}(n)$ be the solution
of \refeq{define_A} corresponding to $p_{(n)}$. We also define a 
three-tensor
\be
    B_{ijk}(n) \equiv c_{ijk1} + p_{(n)} c_{ijk2}, \hspace{2em} n=1,2,3.
\end{equation}
Notice that $p_{(n)}$, $\vect{A}(n)$ and $\vect{B}(n)$ depend
on the line direction of the dislocation and the slip plane, but
not on the Burgers vector, so they are applicable to 
the perfect dislocation and to the two partials.

We now define, depending on the Burgers vector $\vb$, three complex
numbers $D(n)$ by the six equations
\bea
     \Re\left[ \sumn A_k(n) D(n) \right] &= b_k     \notag \\
     \Re\left[ \sumn B_{k2l}(n) A_{l}(n) D(n) \right] &= 0,
\end{align}
Where $\Re(z)$ is the real part of $z$.
This does involve the Burgers vector, so we have $D^{(l)}(n)$, 
$D^{(r)}(n)$, and $D^{(f)}(n)$. However the equations are linear
in $\vb$ and $D$, so that we have
\be
   D^{(l)}(n) + D^{(r)}(n) = D^{(f)}(n).
\end{equation}
We now have \cite{sextic}
\be 
  \sigma_{ij}^{(\#)}(x_{(\#)},0,z) = 
     \frac{-1}{2 \pi x_{(\#)}} 
            \Im\left[ \sumn B_{ijk}(n) A_{k}(n) D^{(\#)}(n) \right],
\end{equation} 
where $(\#)$ is $(l)$, $(r)$ or $(f)$, the relevant dislocation
is situated at $x_{(\#)}=0$, and $\Im(z)$ is the imaginary part of $z$.
Since our slip plane is perpendicular to $x_2$, we have
\be
   t^{(\#)}_i = \sigma_{i2}
      = \frac{-1}{2 \pi x_{(\#)}} 
      \Im \left[ \sumn B_{i2k}(n) A_{k}(n) D^{(\#)}(n) \right].
\end{equation}
We now define 
\bea 
    F_l& = \frac{1}{4 \pi} \Im \left[
     \sumn b^{(l)}_i B_{i2k}(n) A_k(n) D^{(l)}(n) \right], \\
    F_r& \mbox{\ similarly},                                      \\
    F_{\subrm{int}}& = \frac{1}{4 \pi} 
      \Im \left[ \sumn  b^{(l)}_i B_{i2k}(n) A_k(n) D^{(r)}(n) \right]  \\
      & = \frac{1}{4 \pi} 
       \Im \left[ \sumn  b^{(r)}_i B_{i2k}(n) A_k(n) D^{(l)}(n) \right],
\end{align}
where the last equality holds because the interaction forces between
the partials must be equal and opposite.
Thus we have
\bea
      \vt^{(l)} \cdot \vb^{(l)} &= - 2 F_l / x_l  \\
      \vt^{(r)} \cdot \vb^{(r)} &= - 2 F_r / x_r  \\
      \vt^{(r)} \cdot \vb^{(l)} &= - 2 F_{\subrm{int}} / x_r  \\
      \vt^{(l)} \cdot \vb^{(r)} &= - 2 F_{\subrm{int}} / x_l.
\end{align}
From \refeq{tau_dot_b} we have
\bea
  \dea / l &=
        - \left(F_l +F_{\subrm{int}}\right)   
              \int_{-R-d+s}^{-R+s} x^{-1} dx
        - \left(F_r + F_{\subrm{int}}\right)  
               \int_{-R-d-s}^{-R-s} x^{-1} dx             \\
        &= \left(F_l +F_{\subrm{int}}\right)
            \log\left( \frac{R+d-s}{R-s} \right)
          + \left(F_r +F_{\subrm{int}}\right)
            \log\left( \frac{R+d+s}{R+s} \right),
               \label{eq:Ea_partial}
\end{align}
which reduces to \refeq{Ea}, as it should, on setting $s$ equal to zero.

The forms for $\deb$ and $\dec$ at equations (\ref{eq:Eb}) and 
(\ref{eq:Ec}) are
directly applicable to the case where the dislocation is split
into partials, so \refeq{Ea_partial} is the only additional
formula needed for the elasticity computation.

\end{document}